\documentclass[fleqn,usenatbib]{mnras}

\usepackage{newtxtext,newtxmath}
\usepackage[T1]{fontenc}

\DeclareRobustCommand{\VAN}[3]{#2}
\let\VANthebibliography\thebibliography
\def\thebibliography{\DeclareRobustCommand{\VAN}[3]{##3}\VANthebibliography}

\usepackage{graphicx}	
\usepackage{amsmath}	
\usepackage[version=4]{mhchem}
\usepackage{gensymb}
\usepackage{chemfig}

\usepackage{textcomp}

\let\oldAA\AA
\renewcommand{\AA}{\text{\normalfont\oldAA}}

\def\orcid#1{\kern .08em\href{https://orcid.org/#1}{\includegraphics[keepaspectratio,width=0.7em]{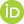}}}

\def\apj{ApJ}

\def\aap{A\&A}
\def\apjl{ApJL}
\def\apjs{ApJS}
\def\mnras{MNRAS}
\def\aj{AJ}
\def\nat{Nature}
\def\grl{GRL}
\def\icarus{Icarus}

\def\aaps{A\&AS}
\def\psj{Planet. Sci. J.}

\title[WD-1856b GCM]{\centering{Modelling the 3D atmospheric structure of the cold Jupiter WD1856+534b orbiting a white dwarf}}

\author[Noti et al.]{
Pascal A. Noti$^{1,2}$ \orcid{0000-0002-8012-3400},
Elspeth K.H. Lee$^{1}$ \orcid{0000-0002-3052-7116},
Daniel Kitzmann$^{2}$ \orcid{0000-0003-4269-3311},
Ryan MacDonald$^{3,4}$\orcid{0000-0003-4816-3469},
\newauthor Sydney Jenkins$^{5}$\orcid{0000-0001-9827-1463},
Arjun Savel$^{6}$ \orcid{0000-0002-2454-768X},
Mary Anne Limbach$^{7}$ \orcid{0000-0002-9521-9798},
Christoph Mordasini$^{2}$ \orcid{0000-0002-1013-2811} \\
$^{1}$Center for Space and Habitability, Universit\"at Bern, Gesellschaftsstrasse 6, CH-3012 Bern, Switzerland \\
$^{2}$Physikalisches Institut, Universit\"at Bern, Sidlerstrasse 5, CH-3012 Bern, Switzerland \\
$^{3}$Department of Astronomy, University of Michigan 1085, Ann Arbor, MI 48109, United States of America\\
$^{4}$School of Physics and Astronomy, University of St. Andrews, North Haugh, St. Andrews, KY16 9 SS, United Kingdom\\
$^{5}$Department of Physics and Kavli Institute for Astrophysics and Space Research, Massachusetts Institute
of Technology, Cambridge,\\ MA 02139, United States of America\\
$^{6}$Department of Astronomy, University of Maryland, College Park, United States of America\\
$^{7}$Department of Astronomy, Cornell University, 404 Space Sciences Building, 14853, Ithaca, NY, United States of America
}

\date{Accepted YY; Received YY; in original form YY}

\pubyear{2025}

\begin{document}
\label{firstpage}
\pagerange{\pageref{firstpage}--\pageref{lastpage}}
\maketitle

\begin{abstract}

 WD-1856b+534b (WD-1856b) is to date the only detected cold Jupiter outside of our Solar System. This cold Jupiter  can provide useful information about the cold giants in our Solar System. Recent JWST observations have targeted WD-1856b, with more scheduled in the near future. To support the interpretation of these observations, we simulated WD-1856b using a three-dimensional (3D) General Circulation Model (GCM) and produced synthetic emission spectra of the planet. We used the Exo-FMS GCM with correlated-k radiative transfer (RT) and mixing-length theory (MLT). In addition, we included abundances of 13 chemical species using the thermochemical kinetic model mini-chem. Because there are substantial uncertainties in the metallicity and internal temperature of WD-1856b, we ran simulations with 1x, 10x, and 100x solar compositions and at low and high internal temperatures (100~K and 500~K). We generated emission spectra and brightness temperature curves with the GCM output using the 3D Monte Carlo radiative-transfer code gCMCRT. Our results suggest larger volume mixing ratios (VMR) of CO and \ce{CO2} with a warmer core at higher metallicity. With a colder core, \ce{H2O} and \ce{CH4} become more relevant and increase to 0.01 VMR at 100x Solar. We suggest possible \ce{H2O} cloud formation in the upper atmosphere in the warm 100x solar case and in all cold cases, which may reduce gas phase \ce{H2O} in the upper atmosphere moderately.
\end{abstract}



\section{Introduction}

The WD-1856b+524b presents a rare opportunity to study cold Jupiters outside of our Solar System. Moreover, \citet{limbach2025thermalemissionconfirmationfrigid} published the emission spectra of WD-1856b. So far, it is the coldest emission spectra measured for an exoplanet, with an equilibrium temperature of 165 K. The second-coldest emission is measured from a temperate super-Jupiter with an equilibrium temperature of 275 K \citet{Matthews2024Natur.633..789M}. This gap in temperature shows how unique WD-1856b is in the known population of exoplanets. Therefore, WD-1856b is the only possibility to compare Jupiter with a similar type of planet.

In the review chapter of \citet{Guillot2023ASPC..534..947G}, the gas giants in our Solar System distinguish themselves by their extraordinary banded structure, spatial variabilities, and strong zonal winds in these bands. These systems of planetary bands are created because of the dominance of the Coriolis force in the momentum balance, the injection of energy and momentum from small scales (eddies and storms) to larger scales (zonal jets), and the conservation of angular momentum and potential vorticity as differentially heated air moves with latitude. These bands are detected in the shallowly measurable troposphere. The deeper structure of the bands remains hidden from spacecrafts' instruments. The bands alternate between prograde and retrograde zonal jets, regions with different temperatures, cloud opacity, and chemical compositions. These seen patterns are thought to be the product of the vertical and meridional circulation cells. The atmospheres of the gas giants in our Solar System can be classified into two groups \citep{Guillot2023ASPC..534..947G}:
\begin{itemize}
    \item Fast rotators ($\sim$ 10 hours), like Jupiter and Saturn, with about 5-8 bands in each hemisphere and a relatively uniform distribution of key condensables away from the equator.
    \item Intermediate rotators ($\sim$ 17 hours), like Uranus and Neptune, with one equatorial retrograde jet, a single prograde jet in each hemisphere, and strong equator-to-pole gradients in condensables.
\end{itemize} 

Hydrogen and Helium dominate cold Jupiters as in our Solar System's Jupiter. There are many other chemical species in more dilute concentrations in the atmosphere of our gas giant, and they occur in both gas and condensate phases \citep{Rensen2023RemS...15..841R}. Studying the abundances of chemical species in the atmosphere of Jupiter is crucial to understand its dynamics, cloud-forming processes \citep{Guillot2020JGRE..12506403G, Guillot2020JGRE..12506404G}, and its formation history \citep{Mousis2009ApJ...696.1348M}. We may expect even more chemical species in different variants of cold gas giants. For instance, WD-1856b has higher incoming stellar radiation that facilitates more chemical reactions on the planet's dayside than in Jupiter.

Our Jupiter is dominated by thick and high-altitude clouds of icy ammonia that block observations of the deeper atmosphere \citep{Rensen2023RemS...15..841R}. If, however, our Jupiter challenges any mission to measure the chemical abundances at higher pressures, we will face harder conditions in determining chemical species in extrasolar cold Jupiters. Nonetheless, the Juno mission's microwave radiometer \citep{Bolton2017Sci...356..821B} managed to hack the clouds by observing at different wavelengths, higher resolutions, and higher latitudes \citep{Bolton2017Sci...356..821B, Grassi2020JGRE..12506206G, Li2020NatAs...4..609LJuno, Visscher2020JGRE..12506526V}. The Juno mission provided intensively studied \ce{CH3} and \ce{H2O} abundances \citep{Bolton2017Sci...356..821B, Li2017GeoRL..44.5317L, Li2020NatAs...4..609LJuno}. These molecules are important carriers of heavier elements in Jupiter's atmosphere. Knowing the abundances helps to understand the dynamics. For instance, gaseous \ce{H2O} prevents dynamic instability across the \ce{H2O} condensation layer that likely has a super-adiabatic temperature gradient at Jupiter's equatorial zone \citep{Li2024Icar..41416028L}. The advection of \ce{CH3} and other chemical species happens in the series of stacked meridional circulation cells (belts), similar to the Ferrel circulation cells on Earth \citep{Ingersoll2017GeoRL..44.7676I, Duer2021GeoRL..4895651D, Fletcher2021JGRE..12606858F}. The gases are transported in opposite directions above and below a transitional layer somewhere between 10$^{5}$ and 10$^{6}$ Pa. This process creates \ce{CH3}-depletion at shallower depths and \ce{CH3}-enrichment at higher pressures. In addition, these measurements allow the study of the formation and precipitation of hail-like particles of dissolved \ce{CH4} on \ce{H2O}-ice or 'mushballs' \citep{Guillot2020JGRE..12506403G, Guillot2020JGRE..12506404G}. Furthermore, Galileo and Juno's measured abundances help us understand the evolution of Jupiter \citep{Helled2014MNRAS.441.2273H, Bosman2019A&A...632L..11B, Oberg2019AJ....158..194O, Howard2023A&A...672L...1H, Howard2023A&A...672A..33H}. Finally, modelling of the chemistry in Jupiter's atmosphere, as in \citet{Barshay1978Icar...33..593B}, \citet{Carlson1987ApJ...322..559C} and in \citet{Fegley1994Icar..110..117F}, complements the measurements and the understanding of Jupiter. For the same reason, we want to support the understanding of WD-1856b's observations.

Regarding the modelling side, \citet{Miguel2022A&A...662A..18M} modelled a T-p profile by extending the \textit{in situ} Galileo measurements with a deep adiabat using the CEPAM code of \citet{Guillot1995A&AS..109..109G}. Their theoretical study and the measurements of the temperature in Jupiter's atmosphere can constrain the temperature structure of WD-1856b.


WD-1856b orbits a cold WD ($T_{\rm WD, eff}$ = 4710 K, log g = 7.902 cm s$^{-2}$) every 1.4 days at a distance of 0.0204 au, resulting in an equilibrium temperature of $T_{\rm P,eq}$ =  163 K \citep{Vanderburg2020Natur.585..363V}. WD-1856b is an ideal object for investigating any orbital disruption caused by the RGB phase \citep{Debes2002ApJ...572..556D, Zuckerman2010ApJ...722..725Z}. WD planetary systems such as WD-1856b are useful indicators into whether planets can survive the RGB phase and remain habitable around stellar remnants \citep{Agol2011ApJ...731L..31A, Fossati2012ApJ...757L..15F, Barnes2013AsBio..13..279B, Loeb2013MNRAS.432L..11L, Kozakis2020ApJ...894L...6K,Becker2023ApJ...945L..24B,Zhan2024ApJ...971..125Z}.

Should WD-1856b originally orbited and formed further out than its current position, WD-1856b might have gone through common-envelope evolution \citep{Lagos2021MNRAS.501..676L, Chamandy2021MNRAS.502L.110C}, migrating inwards and passing through the outer layers of the red giant. After the ejection of the outer layers, WD-1856b continued the migration until its current close orbit. Alternatively, WD-1856b might have interacted gravitationally (high-eccentricity migration) with other planets \citep{Maldonado2021MNRAS.501L..43M} or with nearby stars, such as the red dwarf binary G 229–20,\citep{Munoz2020ApJ...904L...3M, OConnor2021MNRAS.501..507O, Stephan2021ApJ...922....4S} after the RGB phase. The orbit of WD-1856b might have been highly eccentric, and tidally circularized over time.

\citet{Vanderburg2020Natur.585..363V} identified WD-1856b in data from NASA’s Transiting Exoplanet Survey Satellite (TESS) mission \citep{ricker2014} in August 2019. In October 2019, \citet{Vanderburg2020Natur.585..363V} observed other transits with three small privately operated telescopes and another transit with two larger telescopes, the Telescopio Carlos Sánchez and Gran Telescopio Canarias. In December 2019, \citet{Vanderburg2020Natur.585..363V} observed another transit of WD-1856b at wavelengths between 4 and 5 {\textmu}m with NASA’s Spitzer Space Telescope. 
Recently, JWST measured a transmission spectrum of WD-1856b using the NIRSpec/PRISM in BOTS (Bright Object Time-Series) mode \citep{MacDonald2021jwst.prop.2358M}. This mission will provide the first emission spectrum of a planetary companion orbiting a WD.

Following up on the NIRSpec observations, \citet{limbach2025thermalemissionconfirmationfrigid} used the JWST Mid-Infrared Instrument (MIRI) photometry modes to measure the amount of infrared excess produced by the planetary companion.
\citet{limbach2025thermalemissionconfirmationfrigid} updated several system and planetary parameters ($T_{\rm WD,eff}$ = 4920 K, $R_{\star}$ = 0.0121 $R_{\odot}$, a = 0.02085 au). Resulting in a similar $T_{\rm P, eq}$ of 165 K. Their calculations of the effective temperature of the planet based on the atmospheric retrieval lead to $T_{\rm eff}$ = 184 K.
This places it at a much cooler temperature regime than the white dwarf-brown dwarf (WD-BD) population \citep[e.g.][]{Casewell2015}, making it more similar to Solar system gas giants, with their more modest irradiation. 

Several studies have used GCMs to simulate atmospheres across the large exoplanet parameter space, from warm/temperate sub-Neptunes of K2-18b \citep[e.g.][]{showman2015, Rauscher2017ApJ...846...69R, Charnay2021A&A...646A.171C, christie2022}, GJ-1214b \citep[e.g.][]{Kempton2023Natur.620...67K, christie2022}, or hot Jupiters \citep[e.g.][]{showman2009, Ohno2019ApJ...874....1O}. This study contributes as a first GCM simulation of a cold Jupiter orbiting a WD.
We perform cloud-free GCM simulations of the cold Jupiter WD-1856b in order to investigate the atmospheric dynamics, chemistry and 3D thermal structures of this enigmatic object.
Following \citet{Vanderburg2020Natur.585..363V}, we primarily assume Solar metallicity, but also include metal enhanced 10x and 100x Solar scenarios.

Additionally, we compare the T-p profiles of the GCM results to the radiative-convective-equilibrium (RCE) 1D atmosphere models ATMO2020 \citep{Phillips2020A&A...637A..38P} and of the Sonora Bobcat model  \citep{Marley2021ApJ...920...85M} at Solar metallicity. The ATMO2020 and Sonora Bobact models are new generations of substellar atmosphere and evolution models. They were made to study L-, T-, and Y-type brown dwarfs and self-luminous extrasolar planets.

In Section \ref{sec:setup_WD}, we present the setup of the GCM simulation for WD-1856b.
In Section \ref{sec:results_WD}, we present the results of the GCM simulation and compare them with available 1D RCEs ATMO2020 and Sonora Bobact. 
In Section \ref{sec:post-processing_WD}, we produce synthetic emission spectra and brightness temperature curves generated by post-processing the GCM outputs with the gCMCRT code.
Section \ref{sec:disc_WD} contains a discussion of our results. Finally, Section \ref{sec:conc_WD} deals with the summary and conclusions of this study.

\section{Simulation setup}
\label{sec:setup_WD}

\begin{table*}
\centering
\caption{Adopted Exo-FMS GCM simulation parameters for the WD-1856b GCM. We use a cubed-sphere resolution of C48 ($\approx$ 192 $\times$ 96 in longitude $\times$ latitude).}
\label{tab:GCM_parameters}
\begin{tabular}{c c c ll}  \hline \hline
 Symbol & Value  & Unit & Description & Reference and notes \\ \hline
 $T_{\rm int}$ & 100, 500 & \lbrack K\rbrack & Internal temperature & - \\
 $p_{\rm 0}$ & 10$^{8}$ &  \lbrack Pa\rbrack & Reference surface pressure & - \\
 $p_{\rm up}$ & 0.1 &  \lbrack Pa\rbrack & Upper boundary pressure & - \\
 $c_{\rm p}$ & 12'053, 10'681, 4'926&  \lbrack J K$^{-1}$ kg$^{-1}$\rbrack & Specific heat capacity in the dynamical core & calculated based on the metallicity cases \\
 $R_{\rm d}$ &  3'568, 3'149, 1'394&  \lbrack J K$^{-1}$ kg$^{-1}$\rbrack  & Specific gas constant in the dynamical core & calculated based on the metallicity cases \\
 $\kappa$ & $R_d/c_p$ & -  & Adiabatic coefficient & - \\
 $g_{\rm p}$ & 221.0 & \lbrack m s$^{-2}$\rbrack & Surface gravity & - \\
 $R_{\rm p}$ &  6.5186385$\cdot$10$^{7}$  & \lbrack m\rbrack & Planetary radius & \citet{Vanderburg2020Natur.585..363V}\\
 
 $M_{\rm p}$ & 7.4 & \lbrack Jupiter mass\rbrack & Planetary mass & from JWST mission \citep{MacDonald2021jwst.prop.2358M}\\
 $\Omega_{\rm p}$ &  5.165$\cdot$10$^{-5}$ & \lbrack rad s$^{-1}$\rbrack & Rotation rate & \citet{Vanderburg2020Natur.585..363V} \\
 $\left[{\rm M/H}\right]$ & 0,1,2 & - & $\log10$ Solar metallicity & - \\

 $a$ &  0.02016  & \lbrack AU\rbrack & semi-major axis & from JWST mission \citep{MacDonald2021jwst.prop.2358M}\\
 $M_{\star}$ &  0.518  & \lbrack Solar mass\rbrack & Stellar mass & \citet{Vanderburg2020Natur.585..363V}\\
 $R_{\star}$ & 0.0131  & \lbrack Solar radius\rbrack & Stellar radius & \citet{Vanderburg2020Natur.585..363V}\\
  $T_{\rm \star, eff}$ & 4710 & \lbrack K\rbrack & Stellar effective temperature & \citet{Vanderburg2020Natur.585..363V} \\

 $\alpha$ & 1 & - & MLT scale parameter & \citet{Lee2022ApJ...929..180L_gCMCRT} \\
 $\beta$ & 2.2 & - & Overshooting parameter & \citet{Lee2022ApJ...929..180L_gCMCRT} \\
 $\tau_{\rm drag}$ & 10$^{6}$ & \lbrack s\rbrack & Basal drag timescale & - \\
 $p_{\rm \tau,bot}$  & $1\cdot10^{7}$ & \lbrack Pa\rbrack  & Rayleigh pressure threshold & -\\
 $\Delta$ t$_{\rm hyd}$ & 30  & \lbrack s\rbrack & Hydrodynamic time step & - \\
 $\Delta$ t$_{\rm rad}$  & 30 & \lbrack s\rbrack & Radiative time step & - \\
 $\Delta$ t$_{\rm MLT}$ &  0.5 & \lbrack s\rbrack & Mixing length theory time step & \citet{Lee2022ApJ...929..180L_gCMCRT} \\ 
 $\Delta$ t$_{\rm chem}$  & 3'600 & \lbrack s\rbrack & Mini-chem time-step & \citet{Lee2023MNRAS.523.4477L} \\
 $N_{\rm v}$ & 60  & - & Vertical resolution & - \\
 $D_{\rm d\!i\!v,4}$ & 0.16  & - & $\mathcal{O}$(4) Divergence dampening coefficient & - \\
 $D_{\rm h\!y\!p,v}$ & $0.001$ & - & $\mathcal{O}$(4) Vertical hyperdiffusion coefficient & - \\
 $t_{\rm tot}$ & 2'100 & \lbrack Earth days\rbrack & Run length & - \\
\hline
\end{tabular}
\end{table*}

We use the Exo-FMS GCM in the hot Jupiter configuration \citep[e.g.][]{lee2021}, similar to that previously used for WD-BD GCM simulations \citep[e.g.][]{Lee2024MNRAS.529.2686L}.
For the GCM parameterisation, we took values derived from the \citet{Vanderburg2020Natur.585..363V} observations and from the JWST mission \citep{MacDonald2021jwst.prop.2358M}, resulting in the values presented in Table \ref{tab:GCM_parameters}.
We assumed that the planet is on a tidally locked, circular orbit to the WD, with the rotation period of the planet equal to the orbital period. The timescale for the tidal spin-down of the planet, $\tau$ [s], is defined \citep{Guillot1996ApJ...459L..35G} as
\begin{equation}
\tau = Q \Biggl(\frac{R_{\rm p}^{3}}{GM_{\rm p}}\Biggr)\omega_p \Biggl(\frac{M_p}{M_{\star}}\Biggr)^2\Biggl(\frac{D}{R_{\rm p}}\Biggr)^6,
 \label{eq:tidally_spin_down}
\end{equation}
where $Q$ denotes the planets tidal dissipation factor, $R_{\rm p}$ [m] the planetary radius, $G$ [m$^{3}$ kg$^{-1}$ s$^{-2}$] the gravitational constant, $M_{\rm p}$ [kg] the planetary mass, $\omega_p$ [rad s$^{-1}$] the planetary rotation rate, $M_{\star}$ [kg] the stellar mass, and $D$ [m] the small body orbital distance. When we set $Q$ to $1\cdot 10^5$ (similar to Jupiter) and the other parameters as in Table \ref{tab:GCM_parameters}, then we get $\tau \sim 9.4707 \cdot 10^4$ Earth years.  Therefore, the tidal spin-down of the planet is far below the age of the system of $7.930$ Gyr \citep{Vanderburg2020Natur.585..363V}, and we assume synchronous rotation. This is the case even if we consider an uncertainty of total system age between 7.4 and 10 Gyr \citep{limbach2025thermalemissionconfirmationfrigid}.

The internal temperature, $T_{\rm int}$ [K], is unknown and uncertain from the JWST measurements. To estimate the $T_{\rm int}$ of the planet, we use an approach for brown dwarfs \citep[see equation 2.59 in][]{Burrows1993RvMP...65..301B} applied to Jupiter-like planets as
\begin{equation}
L_{\rm p} \sim L_{\rm Jup,int}\Biggl(\frac{M_{\rm p}}{M_{\rm Jup}}\Biggr)^2\Biggl(\frac{5{\rm Gyr}}{t}\Biggr),
 \label{eq:internal_temperature}
\end{equation}
where $L_{\rm p}$ [J s$^{-1}$] is the luminosity of the planet, $L_{\rm Jup,int}$ [J s$^{-1}$] the luminosity of Jupiter's internal temperature, $M_{\rm p}$ [kg] the mass of the planet,  $M_{\rm Jup}$ [kg] the mass of the planet, and $t$ [yr] the age of the system. When we set parameters as in Table \ref{tab:GCM_parameters}, we get through $L=4\pi R^2 \sigma T^4_{\rm eq}$ that $T_{\rm int}\sim$ 251 K. Considering the uncertainty of total system ages with 7.4 to 10 Gyr as in \citet{limbach2025thermalemissionconfirmationfrigid}, we get $T_{\rm int}\sim$ 255 to 236 K. If we additionally reduce the planetary mass to 5.2 $M_{\rm Jup}$, we get $T_{\rm int}\sim$ 214 to 198 K. Since the system could have gone through a common envelope RGB phase, the red giant might have slowed down the cooling or even warmed up the planet. Since the planetary mass is clearly below 13 $M_{\rm Jup}$, we expect no deuterium-burning in WD-1856b \citep{Spiegel2011ApJ...727...57S}. Therefore, we assume a Y-dwarf-like $T_{\rm int}$ = 500 K and a Jupiter-like $T_{\rm int}$ = 100 K as an upper and lower bound to investigate any atmospheric differences. Hopefully, future observations and simulations can provide more clarity regarding $T_{\rm int}$. Because the RGB phase may have enriched the metallicity of the atmosphere, we include 10x and 100x Solar enriched atmosphere scenarios as well as the fiducial 1x models.

We use a cubed-sphere resolution of C48 ($\approx$ 192 $\times$ 96 in longitude $\times$ latitude), which is a suitable spatial resolution given the slower rotation rate of the planet \citep{mayne2019}.
To stabilize the simulations in the deep atmosphere, we included a linear Rayleigh `basal' drag similar to \citet{Tan&Komacek2019} and \cite{Lee2024MNRAS.529.2686L}, specified in Table \ref{tab:GCM_parameters}.

\subsection{Mini-chem}

We couple the miniature chemical kinetics model mini-chem \citep{Tsai2022, Lee2023minichemicalscheme} and the Exo-FMS GCM.
This scheme reduces the number of chemical species to 13 (OH, \ce{H2}, \ce{H2O}, H, CO, \ce{CO2}, O, \ce{CH4}, \ce{C2H2}, \ce{NH3}, \ce{N2}, HCN and He), the number of reactions to 10 and has been extensively used in GCM modelling of sub-stellar objects \citep[e.g.][]{Lee2023MNRAS.523.4477L, Lee2024MNRAS.529.2686L}. The mini-chem scheme does not include photochemistry, but any change in the abundances is accounted for in the radiative transfer scheme through the gas phase opacity during the GCM simulation.

\subsection{Mixing length theory}
We included the same mixing length theory (MLT) scheme as in \citet{Lee2024MNRAS.529.2686L}, where MLT is used to adjust the adiabat in convective regions in a time-dependent manner. 
The estimated eddy diffusion coefficient, $K_{\rm zz}$ [cm$^{2}$ s$^{-1}$], is then used in a vertical tracer diffusion scheme, emulating the mixing of tracers from convective motions.

\subsection{Radiative-transfer}

We used a white dwarf stellar model at the WD1856+534 parameters given in \citet{Vanderburg2020Natur.585..363V} as the input stellar irradiation flux to the GCM (see stellar constants in Table \ref{tab:stellar_fluxes}).
The opacities are mixed on-the-fly inside the GCM using the adaptive equivalent extinction method from \citet{amundsen2016,Amundsen2017}.  
To calculate both the shortwave and longwave fluxes, we used the plane-parallel \citet{toon1989} two-stream source function method in the model, which is well used in GCM \citep[e.g.][]{Roman2019ApJ...872....1R, lee2021, Teinturier2024A&A...683A.231T} and 1D models \citep[e.g.][]{Marley2021ApJ...920...85M} for exoplanets and brown dwarf atmospheres.

\begin{table}
\centering
\caption{Stellar constants in each short-wave band. They were computed with the stellar fluxes by the relation $(R_{\star}/a)^2 S_{\star}$. The sum of the stellar fluxes is 253.637 Wm$^{-2}$.}
\label{tab:stellar_fluxes}
\begin{tabular}{c c c}  \hline \hline
lower band limit $\lbrack \mu m \rbrack$ & upper band limit $\lbrack \mu m \rbrack$  & $S_0$ $\lbrack Wm^{-2} \rbrack$ \\ \hline
 0.26 & 0.42 & 13.289 \\
 0.42 & 0.61 & 53.621 \\
 0.61 & 0.85 & 64.149 \\
 0.85 & 1.32 & 67.566 \\
 1.32 & 2.02 & 35.817 \\
 2.02 & 2.5 & 7.710 \\ 
 2.5 & 3.5 & 6.918 \\ 
 3.5 & 4.4 & 2.232 \\ 
 4.4 & 8.7 & 2.336 \\ 
 8.7 & 20.0 & 0.0 \\  
 20.0 & 324.68 & 0.0 \\ 
\hline
\end{tabular}
\end{table}

\subsection{Gas phase opacities}

We used a correlated-k approach with 11 bands as in \citet{Kataria2013} and with the adaptive equivalent extinction method detailed in \citet{Amundsen2017} to mix the opacity tables on the fly during the simulation.
This allows the varying VMR of the gas species produced by mini-chem to be fed-back into the temperature structure of the simulation.
For the line opacity, we included species that are present in mini-chem, OH \citep{Hargreaves2019}, \ce{H2O} \citep{Polyansky2018}, CO \citep{Li2015}, \ce{CO2} \citep{Yurchenko2020}, \ce{CH4} \citep{Hargreaves2020}, \ce{C2H2} \citep{Chubb2020}, \ce{NH3} \citep{Coles2019}, HCN \citep{Harris2006}, where the citation for each species denotes the source of the line-list used to produce the correlated-k opacity tables.
We assumed that strongly optical wavelength absorbing species such as TiO, VO and Fe are condensed and rained out of the atmosphere and do not contribute as a gas phase opacity source.
In addition, due to the coldness of WD-1856b, we assumed that Na and K are also rained out and not present in the photosphere.
We included collision-induced absorption from \ce{H2}-\ce{H2}, \ce{H2}-He, \ce{H2}-H and He-H pairs, taking the data from \citet{Karman2019}, as well as Rayleigh scattering from \ce{H2}, He and H.
For the duration of 500 days, we mixed a pre-mixed correlated-k table assuming CE at 1x, 10x and 100x Solar composition \citep{Asplund2021A&A...653A.141A} into the GCM.

\subsection{Specific heat capacity}

We computed the non-constant specific heat capacity given the local VMR of chemical species like in \citet{lee2022}. We used the non-constant specific heat capacity based on the local temperature \citep[JANAF tables][]{Chase1986jtt..book.....C} to calculate the local RT heating rates. This variable specific heat capacity was also used in the MLT scheme. However, we note a globally constant specific heat capacity was still used in the GCM dynamical core.

\section{GCM results}
\label{sec:results_WD}

\begin{figure*}
        \includegraphics[width=0.95\textwidth]{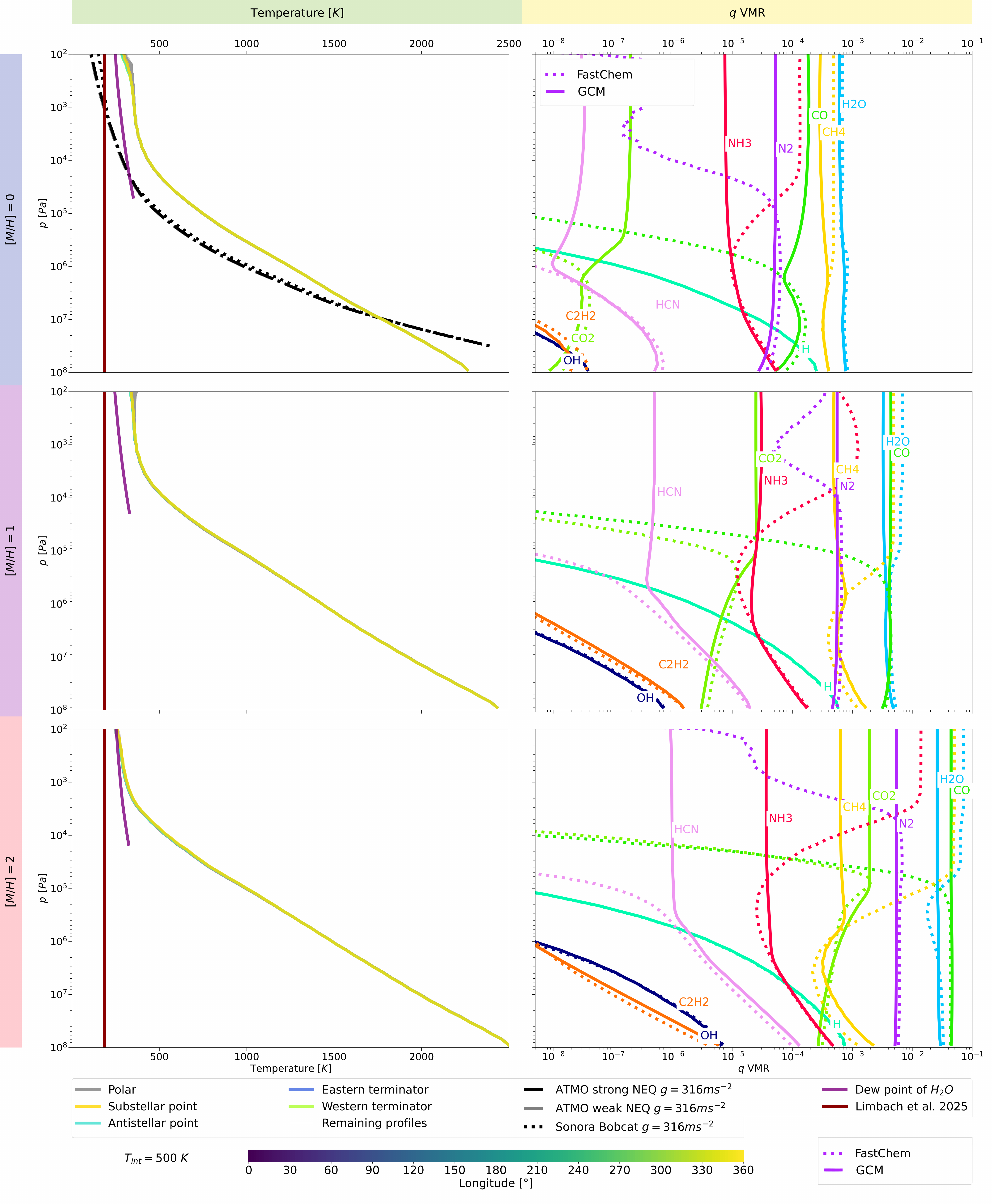}
    \caption{T-p profiles (left column) of WD-1856b and horizontally averaged chemical abundances (right column) produced by the GCM in the 1x, 10x, and 100x Solar cases and at $T_{\rm int}$ = 500 K. For the T-p profiles (left column), the coloured lines indicate vertical profiles along the equator and the colourbar indicate its coordinates. The dark grey lines show the T-p profiles at latitudes 87°N and 87°S. The bold coloured lines represent profiles at the western and eastern terminators, and sub-stellar and anti-stellar points. The lighter grey lines represent all other T-p profiles. The black dotted and dashed lines show the T-p profiles of 1D atmosphere model ATMO2020 with weak and strong non-equilibrium chemistry, $T_{\rm int}$ = 500 K, $g_{\rm p}$ = 100 and 316 m s$^{-2}$ by \citet{Phillips2020A&A...637A..38P} and Sonora Bobcat by \citet{Marley2021ApJ...920...85M} with $T_{\rm int}$ = 500 K $g_{\rm p}$ = 178 and 316 m s$^{-2}$. Moreover, the purple line denotes the the dew point of \ce{H2O}. Additionally, the dark red line indicates the retrieved temperature from the observation of JWST's MIRI by \citet{limbach2025thermalemissionconfirmationfrigid}. The averaged abundances of OH, \ce{H2}, \ce{H2O}, H, CO, \ce{CO2}, O, \ce{CH4}, \ce{C2H2}, \ce{NH3}, \ce{N2}, HCN and He produced by the GCM are shown in the right column. Additionally, the abundances are labeled in different colours and by the labels close to the lines. Finally, the coloured dotted lines (right column) indicate the abundances generated by FASTCHEM 2 \citep{Stock2022MNRAS.517.4070S} on the basis of the mean horizontal T-p profile.}
    \label{fig:temp_q}
\end{figure*}

Figure \ref{fig:temp_q} presents the T-p profiles and horizontally averaged chemical abundances for the $T_{\rm int}$ = 500 K. The temperatures in all cases with the $T_{\rm int}$ = 500 K are horizontally similar. Therefore, WD-1856b with $T_{\rm int}$ = 500 K is in the weak temperature gradient (WTG) limit, presumably due to weak irradiation and slow rotation. Several researchers \citep{Charney1963JAtS...20..607C, Sobel2001JAtS...58.3650S,Pierrehumbert2016RSPSA.47260107P} investigated the WTG behaviour (see Appendix \ref{app:WTG}) that explains temperature gradients on a rotating sphere in the shallow-water theory. This is a suitable approach for dynamics that were computed with the primitive equation set. Since the WTG parameter expresses the competition between heating and rotational effects, we expect similar WTG parameters when $T_{\rm int}$ is the same. However, there are significant differences. We assume that the differences in the WTG parameters arise from radiative feedback from varying chemical abundances. This leads to different dynamics and temperature patterns. The WTG parameter, $\Lambda$, is clearly above 1 for the lower-metallicity cases (see Table \ref{tab:WTG_parameters}).

A change from 1x to 10x Solar composition leads to a shift of the deep adiabat to higher temperatures and a slight cooling of the isotherm in the upper atmosphere. Even higher metallicity leads to much stronger cooling in the upper atmosphere. The stronger cooling brings the local temperatures closer to the dew point of \ce{H2O}. Hence, the 100x Solar case is in the range of water cloud formation in the upper atmosphere. Moreover, there are two major vertical regions with different lapse rates in the deep atmosphere. These two major deep adiabatic regions are separated by less stable layers. These less stable layers increase in extent and instability the higher the metallicity is set.

\begin{table}
\centering
\caption{The WTG parameter, $\Lambda$, computed from the GCM simulations of WD-1856b for the 1x, 10x, and 100x Solar cases at $T_{\rm int}$ = 100 and 500 K. We listed the WTG parameter for each case according the equations presetned in the Appendix \ref{app:WTG}).}
\label{tab:WTG_parameters}
\begin{tabular}{c c c}  \hline \hline
 $T_{\rm int}$ & [M/H]  & $\Lambda$  \\ \hline
 500 & 0 & 1.47 \\
 500 & 1 & 1.46 \\
 500 & 2 & 0.90 \\
 100 & 0 & 0.92 \\
 100 & 1 & 0.93 \\
 100 & 2 & 0.65 \\ 
\hline
\end{tabular}
\end{table}

In Figure \ref{fig:temp_q}, we also compare the T-p profiles of the GCM results to the radiative-convective-equilibrium (RCE) 1D atmosphere models ATMO2020 \citep{Phillips2020A&A...637A..38P} and of the Sonora Bobcat model  \citep{Marley2021ApJ...920...85M} at Solar metallicity.
The RCE models are up to $\sim$ 100 to 200 K cooler at pressures $p\leq 1\cdot10^7$ Pa. At higher pressures, these models predict higher temperatures. In addition, the ATMO2020 and Sonora Bobcat model agree well on temperature predictions, but not with the GCM results. Their deep adiabatic lapse rate is shallower than the lapse rate predicted by the GCM.

Regarding mini-chem results, most chemical abundances increase about 1 or 2 magnitudes if the metallicity is set 1 magnitude higher. \ce{CH4} and \ce{NH3} become only slightly more abundant at higher metallicity. A change from 1x to 10x Solar composition increases the abundances of OH and \ce{C2H2} by 1 magnitude, but the change to 100x Solar composition leads to a minimal increase in the abundances of OH and \ce{C2H2}.

In the deep atmosphere, the mini-chem scheme (disequilibrium chemistry) in the GCM and the post-processing with FastChem 2 (equilibrium chemistry) produce similar abundances for the very most simulated chemical species. In the upper atmosphere, higher metallicities lead to higher divergence between both models. At Solar composition, both models produced relatively similar \ce{H2O} and \ce{CH4} abundances in the upper atmosphere. However, the abundances of the other chemical species differ by several magnitudes.
In the upper atmosphere, \ce{NH3}, \ce{CH4} and \ce{H2O} become more abundant in FastChem 2 with higher metallicity, whereas the other chemical species vanish. Especially, \ce{CH4} experiences the largest increase, by about 2 dex, at 100x Solar composition.

\begin{figure*}
        \includegraphics[width=0.95\textwidth]{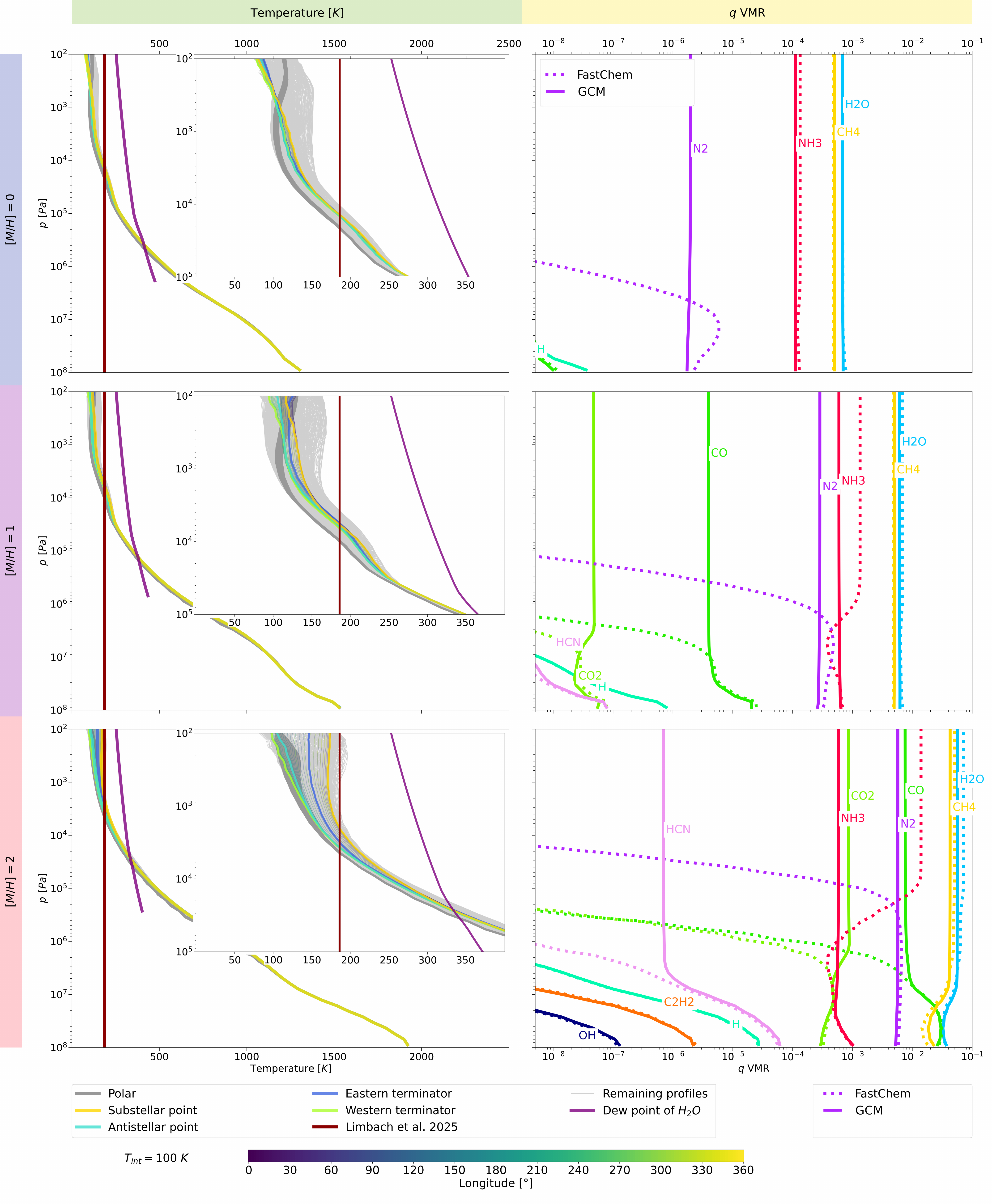}
    \caption{T-p profiles (left column) of WD-1856b and horizontally averaged chemical abundances (right column) produced by the GCM in the 1x, 10x, and 100x Solar cases and at $T_{\rm int}$ = 100 K. For the T-p profiles (left column), the coloured lines indicate vertical profiles along the equator and the colourbar indicate its coordinates. The dark grey lines show the T-p profiles at latitudes 87°N and 87°S. The bold coloured lines represent profiles at the western and eastern terminators, and sub-stellar and anti-stellar points. The lighter grey lines represent all other T-p profiles. Moreover, the purple line denotes the the dew point of \ce{H2O}. Additionally, the dark red line indicates the retrieved temperature from the observation of JWST's MIRI by \citet{limbach2025thermalemissionconfirmationfrigid}. The averaged abundances (right column) of OH, \ce{H2}, \ce{H2O}, H, CO, \ce{CO2}, O, \ce{CH4}, \ce{C2H2}, \ce{NH3}, \ce{N2}, HCN and He produced by the GCM. Additionally, the abundances are labeled in different colours and the labels close to the lines. Finally, the coloured dotted lines (right column) indicate the abundances generated by FASTCHEM 2 \citep{Stock2022MNRAS.517.4070S} on the basis of the horizontally mean T-p profile.}
    \label{fig:temp_q_Tin100}
\end{figure*}

Figure \ref{fig:temp_q_Tin100} presents the T-p profiles and horizontally averaged chemical abundances for the $T_{\rm int}$ = 100 K. The temperatures on the same pressure level become more discrepant in the colder cases. The WTG behaviour is less pronounced, which is supported by the lower WTG parameters (see Table \ref{tab:WTG_parameters}). Similarly as in the warmer cases, higher metallicities shift the deep adiabat to higher temperatures, but there is no trend in the isotherm in the upper atmosphere. The cooler $T_{\rm int}$ leads to local temperatures below the dew point of \ce{H2O} in all metallicity settings. Therefore, we expect that the vertical extent of the water and ice clouds decreases with higher metallicity, and that cloud base becomes shallower. Moreover, the major vertical regions with different lapse rates in the deep atmosphere become more striking and increase to 3 regions.

At 1x to 10x Solar composition, \ce{H2O}, \ce{CH4}, \ce{NH3}, and \ce{N2} are the most abundant chemical species apart from the bulk of \ce{H2} and He abundances.
Higher metallicities lead to a drastic increase in the abundances of \ce{CO} and \ce{CO2} (by nearly 3.5 and 4.5 magnitudes). At 100x Solar composition, \ce{CO} become more abundant than \ce{N2} and \ce{NH3}, and \ce{CO2} more than \ce{NH3}, respectively. \ce{CH4} and \ce{H2O} increase by about one dex for each 10 times higher metallicity. In this way, the 100x Solar case with $T_{\rm int}$ = 100 K evolves abundances of \ce{CH4} and \ce{H2O} of a few percentages in the upper atmosphere. Therefore, we expect the thickest water clouds with higher metallicity and with lower $T_{\rm int}$.

\begin{figure*}
        \includegraphics[width=0.95\textwidth]{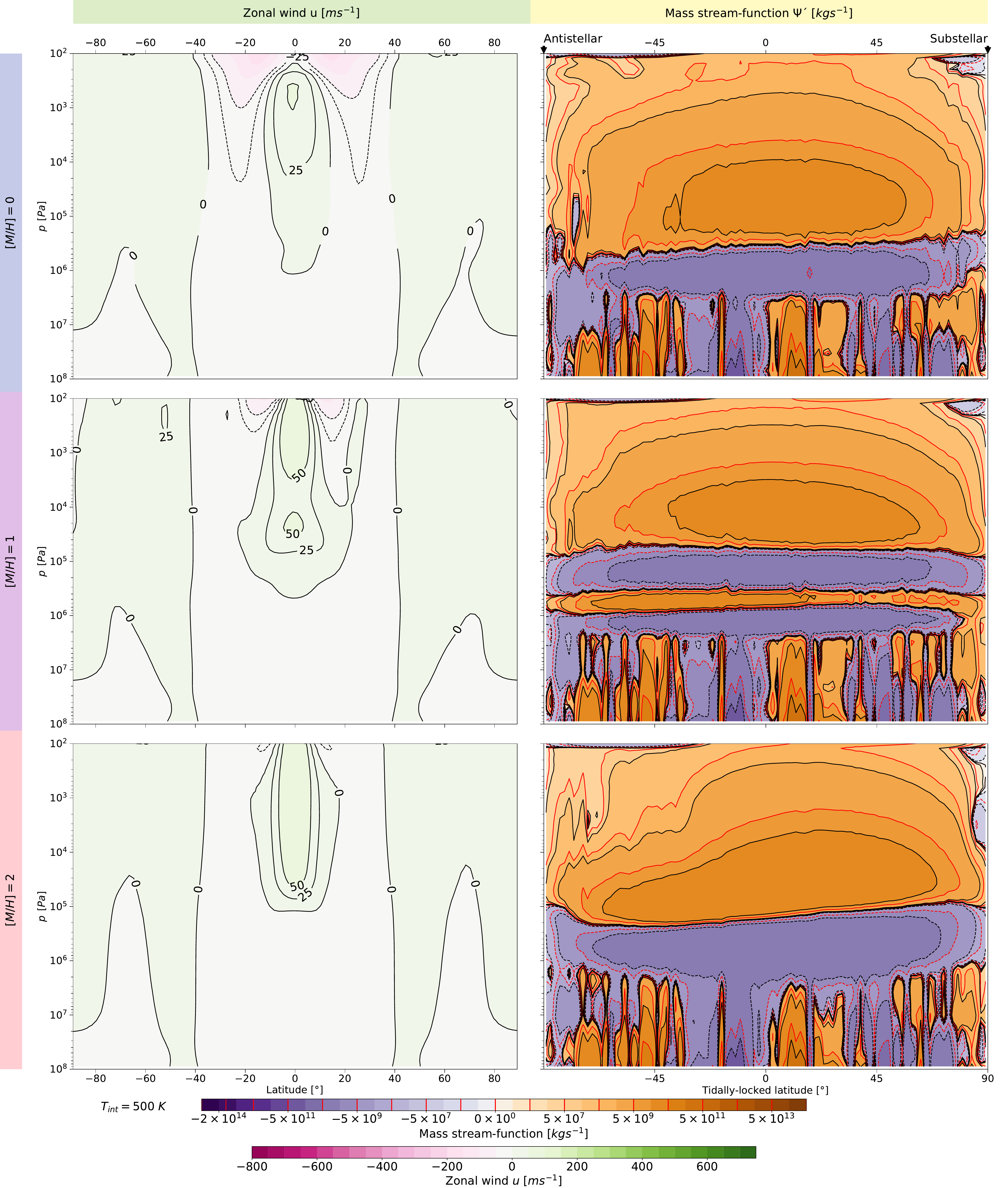}
    \caption{Zonal mean wind and overturning circulation depicted by the mass stream-function $\Psi '$ (second column, tidally-locked coordinates) of WD-1856b for the 1x, 10x, and 100x Solar cases at $T_{\rm int}$ = 500 K. The stream-function shows anti-clockwise and clockwise circulations in orange and in purple, respectively.}
    \label{fig:U_overturning}
\end{figure*}

Figure \ref{fig:U_overturning} shows the zonal wind and the overturning circulation for the $T_{\rm int}$ = 500 K. The westward jet at 100 Pa weakens with higher metallicity. On the other hand, the eastward superrotating jet strengthens and enlarges at higher pressures.

The overturning circulation cell in the upper atmosphere (at $\le 1 \cdot10^5$ Pa) weakens at moderate metallicity and strengthens at high metallicity (see Table \ref{tab:total_mass_overturning}). Similar trends are seen in the underlying circulation cells. Stronger overturning circulation in the 100x solar case leads to a greater heat redistribution that affects the T-p profile and chemistry. The quenching levels lay below the upper circulation cell, except for \ce{CO2} in the 10x and 100x solar case. The increase in the strength of the overturning circulation cell in the upper atmosphere is only aligned with the decreasing WTG parameters in the 100x solar case. The lower metallicity cases contradict an alignment of the circulation strength with the WTG parameters.

\begin{table}
\centering
\caption{The summed overturning mass rate in the upper atmosphere and the circulation depth of the upper most overturning circulation cell. We assessed the circulation depth where the change of the horizontally averaged mass stream function to negative values happens.}
\label{tab:total_mass_overturning}
\begin{tabular}{c c c c}  \hline \hline
 $T_{\rm int}$ & [M/H]  & Overturning rate $\lbrack kgs^{-1} \rbrack$  & Circulation depth $\lbrack Pa \rbrack$\\ \hline
 500 & 0 & 1.42$\cdot10^{13}$ & 4.35$\cdot10^{5}$\\
 500 & 1 & 1.2$\cdot10^{13}$  & 6.28$\cdot10^{4}$\\
 500 & 2 & 1.9$\cdot10^{13}$  & 2.1$\cdot10^{5}$\\
 100 & 0 & 5.29$\cdot10^{13}$ & 1.85$\cdot10^{6}$\\
 100 & 1 & 2.05$\cdot10^{13}$ & 2.68$\cdot10^{5}$\\
 100 & 2 & 4.87$\cdot10^{13}$ & 5.54$\cdot10^{5}$\\ 
\hline
\end{tabular}
\end{table}

\begin{figure*}
        \includegraphics[width=0.95\textwidth]{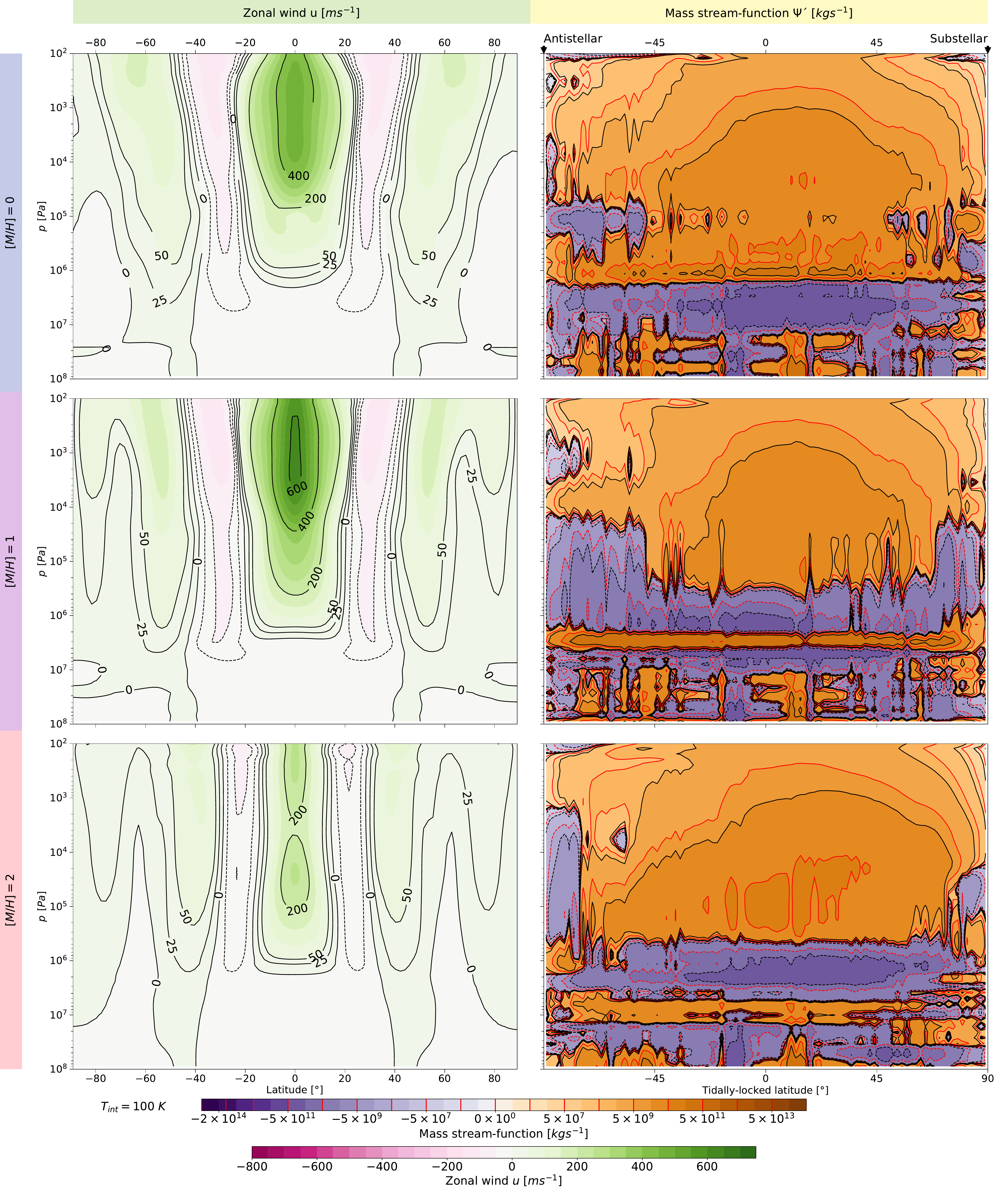}
    \caption{Zonal mean wind and overturning circulation depicted by the mass stream-function $\Psi '$ (second column, tidally-locked coordinates) of WD-1856b for the 1x, 10x, and 100x Solar cases at $T_{\rm int}=100$ K. The stream-function shows anti-clockwise and clockwise circulations in orange and in purple, respectively.}
    \label{fig:U_overturning_Tint100}
\end{figure*}

Figure \ref{fig:U_overturning_Tint100} shows the zonal wind and the overturning circulation for $T_{\rm int}$ = 100 K. The zonal wind weakens at highest metallicity. The strongest eastward superrotating jet is reached in the 10x Solar case. The colder $T_{\rm int}$ cases lead to an increase in the zonal wind.

Similarly, the colder cases evolve stronger overturning circulation cells (see Table \ref{tab:total_mass_overturning}).
Again, the overturning circulation cell in the upper atmosphere (at $\le 1 \cdot10^5$ Pa) cycles the most in the 100x Solar case. The second strongest cell is in the 1x Solar case. Nevertheless, there is uncertainty regarding the evolution of the overturning circulation since the GCM simulated 2100 days. All quenching levels are below the upper circulation cell.

\section{Post-processing}
\label{sec:post-processing_WD}

\begin{figure}
\centering
        \includegraphics[width=0.5\textwidth]{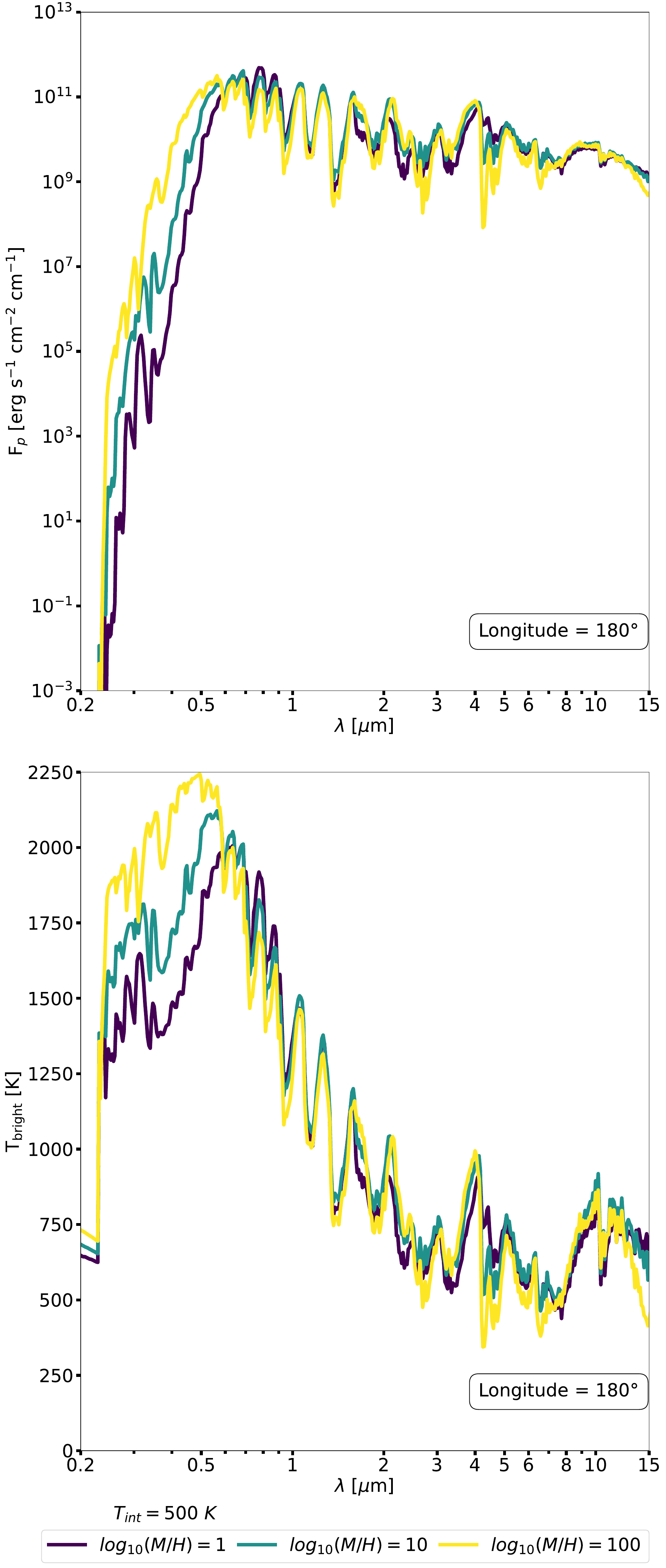}
    \caption{Planetary spectral flux and brightness temperature of WD-1856b at a longitudinal viewing angle of 180$^{\circ}$ (bottom panel) based on the post-processing of the GCM simulations for the 1x, 10x, and 100x Solar cases at $T_{\rm int}$ = 500 K. The line styles denote post-processing.}
    \label{fig:post-processing}
\end{figure}

We used the gCMCRT model \citep{Lee2022ApJ...929..180L_gCMCRT} to post-process the GCM simulation output. 
We produced synthetic spectral phase curves of the model for comparisons to current and future observational missions on WD-1856b (see Figure \ref{fig:post-processing}).

\begin{figure}
\centering
        \includegraphics[width=0.5\textwidth]{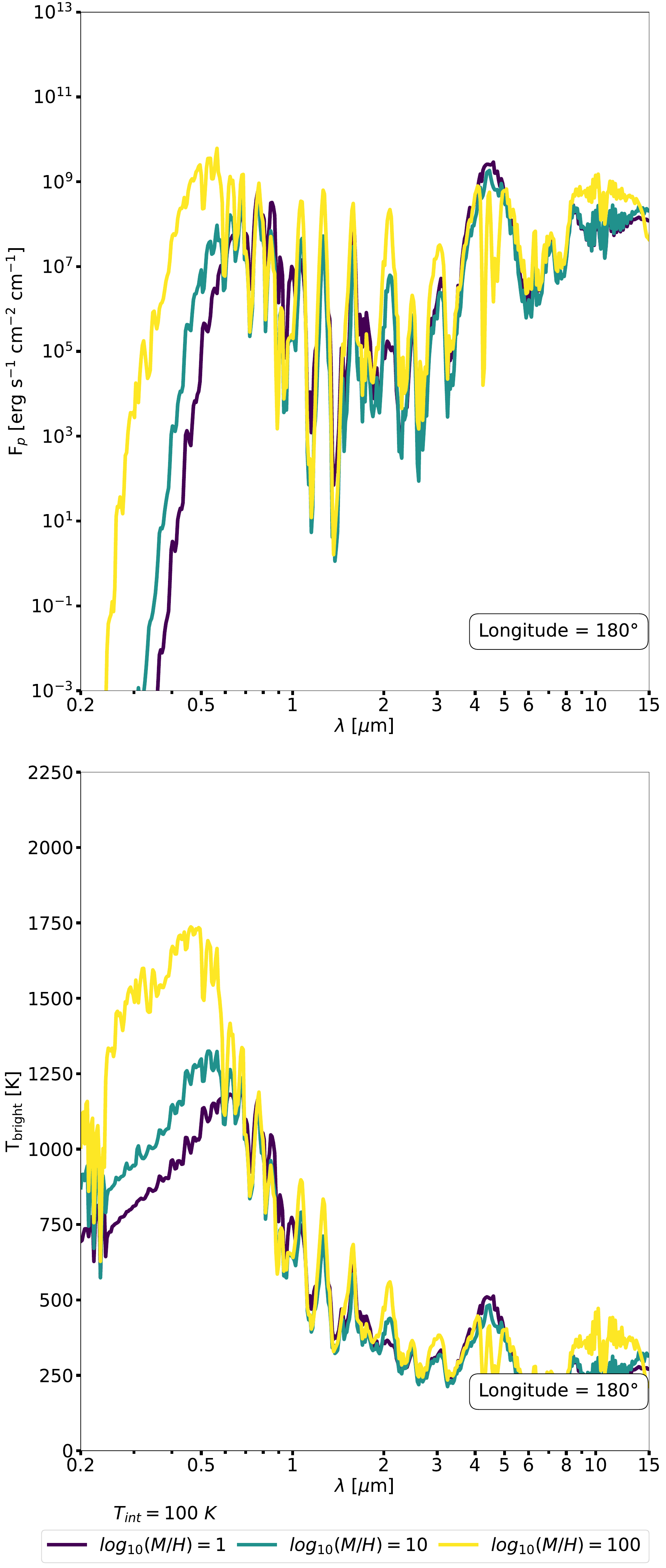}
    \caption{Planetary spectral flux and brightness temperature of WD-1856b at a longitudinal viewing angle of 180$^{\circ}$ (bottom panel) based on the post-processing of the GCM simulations for the 1x, 10x, and 100x Solar cases at $T_{\rm int}$ = 100 K. The line styles denote post-processing.}
    \label{fig:post-processing_Tin100}
\end{figure}

We expect that absorption in the planetary spectral flux in the bands of \ce{CH4} and \ce{H2O} are more pronounced in the colder cases (to some degree \ce{NH3} as well), especially at 100x Solar. If WD-1856b has a higher $T_{\rm int}$, then we expect more absorption characteristics of CO and \ce{CO2}. Therefore, the absorption characteristics of \ce{CH4}, \ce{H2O}, CO and \ce{CO2} in the planetary spectral flux are good indicators for $T_{\rm int}$. CO is always more abundant than \ce{N2} at higher $T_{\rm int}$ and vice versa at low $T_{\rm int}$.
Then, low $T_{\rm int}$ leads to a significant increase in planetary spectral flux at wavelengths of $\lambda > 2$ \textmu m. However, a higher $T_{\rm int}$ lets the flux decrease with higher wavelengths.

At wavelengths of $\lambda < 2$ \textmu m, higher metallicities shift the peak of the planetary spectral flux to smaller wavelengths at high and low $T_{\rm int}$. Additionally, the presence of weak absorption by CO and \ce{CO2} in the low $T_{\rm int}$ case could indicate a 100x Solar composition. Similarly, weak absorption features by \ce{CO2} can help to distinguish a 100x from lower-metallicity compositions. At high $T_{\rm int}$, the abundance ratio of \ce{N2} to \ce{CH4} is a good indicator of the metallicity on WD-1856b. If \ce{CH4} is significantly more abundant than \ce{N2}, the metallicity is near 1x Solar. If the abundances of \ce{N2} and \ce{CH4} are nearly similar, then the metallicity is near 10x Solar. If \ce{N2} is significantly more abundant than \ce{CH4} at high $T_{\rm int}$, the metallicity is close to 100x Solar. Similarly, there is a similar analogy with the abundance ratio of \ce{H2O} to CO at high $T_{\rm int}$. At low $T_{\rm int}$, such analogies occur with the abundance ratio of \ce{N2} to \ce{NH3}.


\section{Discussion}
\label{sec:disc_WD}

Our simulations differ from the 1D atmosphere model simulations of the Sonora Bobcat \citep{Marley2021ApJ...920...85M} and of the ATMO2020 model \citep{Phillips2020A&A...637A..38P} by a higher deep adiabatic lapse rate and higher temperatures in the upper atmosphere (see Section \ref{sec:results_WD}). The higher temperatures in the upper atmosphere are probably caused by the irradiation from the WD, advection, lower gravity, and effective lapse rates in the MLT and in the local RT heating rates in our setup (see Section \ref{sec:setup_WD}). The effective lapse rates, advection, and altered gravity in our setup affects the deep temperature structure as well which explains different deep lapse rates there. All three models used correlated-k RT schemes that differ slightly. Additionally, the GCM redistributes heat through horizontal and vertical advection, which affects the T-p profiles and abundances. Furthermore, the circulation affects the species distribution directly.


The abundances produced by mini-chem \citep{Tsai2022, Lee2023minichemicalscheme} remain vertically almost constantly above each species respective quench pressures, whereas abundances produced by the chemical equilibrium (CE) model FastChem 2 \citep{Stock2022MNRAS.517.4070S} respond to the pressure and temperature strongly like in \citet{Lee2023minichemicalscheme}. All quenching levels are below the upper circulation cell.
In all metallicity cases, quenching occurs at even higher pressures in the cold case.
Our results suggest that any CO and \ce{CO2} features, in an otherwise \ce{CH4} dominated spectrum, present in the JWST data will be a result of this quenching behaviour. Similar behaviour applies to \ce{N2}, CO and \ce{CO2}. CO and \ce{CO2} are quenched at $>$ 1 and $>$ 0.01 ppm abundance, except for the cold Solar case. We see a quenching of \ce{N2} at $>$ 1 ppm abundance.
In addition, from our results, HCN is quenched at $>$ 1 ppm abundance for higher metallicities in the warm and cold case, suggesting it is a good indicator for the metallicity of the atmosphere.
Without quenching, \ce{CH4} in the warm case, \ce{H2O} and \ce{NH3} would be more abundant. Instead, quenching reduces \ce{NH3} and \ce{CH4} up to 2 dex. We can expect \ce{NH3} signatures to be present at 100 $<$ q $<$ 1000 ppm abundance in the cold case, and at 5 $<$ q $<$ 100 ppm abundance in the warm case. Whereas quenching of \ce{H2O} plays a minor role in the cold case, it becomes more relevant at higher metallicity in the warmer case. In a scenario without cloud formation, we expect \ce{H2O} abundances $>$ 500 ppm, and $>$ 1 \% at high metallicity. In addition to quenching, we expect a reduction of \ce{H2O} by cloud formation and rain in the cold case in the upper atmosphere, because the temperatures of all profiles are much lower than the average \ce{H2O} dew point.

Because WD-1856b is the same cold type of planet as the Solar System's Jupiter, a comparison is meaningful. \citet{Rensen2023RemS...15..841R} modelled the chemistry of Jupiter with a Gibbs free energy minimization code (GGCHEM) with a modification for the gas phase and condensation equilibrium chemistry in the deep atmosphere. Their chemical model uses a larger pathway network and more elements than in our study. Additionally, \citet{Rensen2023RemS...15..841R} used enrichment factors with respect to 1x solar for several species based on measurements from Galileo in 2004 in \citet{Wong2004Icar..171..153W} and Juno in 2020 in \citet{Li2020NatAs...4..609LJuno}. Based on the Juno mission, \citet{Rensen2023RemS...15..841R} update the enrichment factors to 2.7 for \ce{H2O} and 2.76 for \ce{NH3}. In addition, they used 3.27 for \ce{CH4} as another enrichment factor based on the Galileo mission in 2004.

Aside from \ce{H2} and He, the most abundant species in Jupiter's upper atmosphere according to \citet{Rensen2023RemS...15..841R} are \ce{H2O}, \ce{CH4} and \ce{NH3}. This descending order is also found in our 1x and 10x cases at $T_{\rm int}=100$ K. In addition, we report slightly higher abundances for \ce{H2O}, \ce{CH4} and \ce{NH3} in our 10x case at $T_{\rm int}=100$ K than \citet{Rensen2023RemS...15..841R}. In comparison, our 1x case has almost 1 dex lower \ce{H2O}, \ce{CH4}, and \ce{NH3} abundances. These species make up the main carriers of O, C, and N in this environment. \ce{H2S} is the next-most abundant species in their main simulation, but it is not included in our scheme. The \ce{N2} abundance in their main simulation is at a similar magnitude in the deep atmosphere, but their main simulation yields more dilute \ce{N2} at lower pressures, whereas our simulation shows minor variation with height. The next less abundant species in their main simulation and in common with our study is \ce{CO}. The descending order in species abundances is the same, but their \ce{CO} abundance lies between our 1x and 10x solar case. There are several species and elements that are not included in our scheme at abundances smaller than 10$^{-5}$.

Across the chemical species, the abundances in \citet{Rensen2023RemS...15..841R} are closer to the cold 10x than the cold 1x solar case. In fact, their enrichment factors are between 2.7 and 3.27. Therefore, our cold 1x solar case should be closer to their results than our 10x solar case. The reasons for the differences likely arise from the chemistry schemes and their settings, because the differences are already significantly large at the lower boundary. It is difficult to assess the mixing in this comparison, because only \ce{N2} shows strikingly different behaviour across the pressure layers in both studies. Moreover, variations due to condensation chemistry in \citet{Rensen2023RemS...15..841R} lead to relatively small changes in the abundances. For example, \ce{NH3} abundance decreases with lower pressures from 10$^{-3.07}$ to 10$^{-3.16}$ in their study. Therefore, we expect similar minor condensational effects in WD-1856b, with only small depletion of \ce{H2O}. As \ce{NH3} cloud formation plays a role in  Jupiter \citep{Rensen2023RemS...15..841R}, \ce{NH3} clouds can be expected in WD-1856b, as well. In the end, the descending order of the chemical species is the same in both studies.

The more recent study of \citet{Li2024Icar..41416028L} assumed that the deep O/H ratio on Jupiter lies between 1.4--8.3x solar, with the more confident estimate at about 4.9x Solar. Therefore, the abundances of \ce{H2O} in our 10x solar case are not so far away from those in Jupiter.

The older study of \citet{Fegley1994Icar..110..117F} resulted in lower abundances of \ce{H2O}, \ce{CH4}, and \ce{NH3} because \citet{Rensen2023RemS...15..841R} used higher enrichment factors for the chemical species based on the Juno data.


In the case of higher $T_{\rm int}$, WD-1856b may have radiatively driven circulation more like a brown dwarf than hot Jupiters due to the small temperature contrasts \citep{Tan2022,TanShowman2021b} and perhaps due to the WTG behaviour \citep{Pierrehumbert2019AnRFM..51..275P}. We cannot determine the WTG behaviour in the 10x solar case, because the summed overturning masse rate in the upper atmosphere is significantly weaker than the 1x solar case, although the $\Lambda$ is similar. In the case of lower $T_{\rm int}$, winds strengthen, and the circulation pattern more resembles that of colder Jupiters \citep{Tan2022}.

We predict the winds and circulation on WD-1856b in the cold 1x and 10x Solar case to be stronger than on Jupiter \citep{Porco2003Sci...299.1541P} and than on Saturn \citep{Garcia2011Icar..215...62G}.  Only the cold 100x solar case (apart from the warm case) has weaker winds than on Saturn. Because WD-1856b receives more radiation from its star, the day-night temperature contrast and temperature gradients are larger, which drives the circulation. However, the primitive equation sets (used in the Exo-FMS GCM) tend to overestimate wind speeds \citep{mayne2019,deitrick2020,Noti2023MNRAS.524.3396N}. Therefore, we can expect weaker winds than we predicted. Additionally, the use of the full equation sets and a higher model resolution may change our predictions to a circulation system that resembles the one on Jupiter or Saturn. Then, the injection of energy and momentum from smaller scales, the missed components of the Coriolis force, and other missed effects in the primitive equation sets could reveal smaller circulation cells (bands). The major jet in our cold case is prograde, as on Jupiter and Saturn. Only the warm 1x and 10x solar cases have few regions with retrograde winds. But our predictions do not include prograde flow at higher latitudes and retrograde flow at the equator like on Uranus \citep{Sromovsky2015Icar..258..192S} and Neptune \citep{Karkoschka2011Icar..215..439K, Carrion2023A&A...674L...3C}. Although the rotation period of 1.4 days is more similar to intermediate rotators, WD-1856b shares more characteristics of fast rotators with respect to the modelled dynamics.

Regarding the temperature structure, \citet{Miguel2022A&A...662A..18M} generated a T-p profile by extending the \textit{in situ} Galileo measurements with a deep adiabat using the CEPAM code of \citet{Guillot1995A&AS..109..109G}. Their model predicts roughly 1'350~K at $p=$ 10$^{8}$ Pa, whereas our model predicts 1'306~K and 1'529~K in the 1x and 10x solar case at $T_{\rm int}=$100 K. At $p=$10$^{5}$ Pa, we predict 273~K and 346~K for both cases, whereas they report roughly 180~K. The large temperature difference at $p=$10$^{5}$ Pa can be explained by the much higher short-wave radiation in WD-1856b. The stellar constant of WD-1856b is 253~Wm$^{-2}$ and surpasses Jupiter's solar constant of 45.9 Wm$^{-2}$ at the maximum and 55.8 Wm$^{-2}$ at the minimum by far. Our 1x solar case is slightly colder at $p=$ 10$^{5}$ Pa, because Jupiter is slightly enriched in heavier elements compared to the solar composition \citep{Wong2004Icar..171..153W, Li2020NatAs...4..609LJuno, Rensen2023RemS...15..841R}. Hence, the radiative feedback results in a different temperature structure. Rhis radiative feedback from more abundant heavier elements therefore explains the higher temperature in our 10x and 100x solar case, as well. So, our colder case is in line with the results in \citet{Miguel2022A&A...662A..18M}.

When comparing the temperature structure with the T$_{\rm eff}$ = 184~K by \citet{limbach2025thermalemissionconfirmationfrigid}, the colder cases evolve T-p profiles in the range of 184~K in the photosphere (see Figure \ref{fig:temp_q_Tin100}). The higher metallicities seem to suit even more. Therefore, our colder cases are in line with \citet{limbach2025thermalemissionconfirmationfrigid}, although a slightly higher $T_{\rm int}$ could match 184~K, as well. More theoretical and observational studies are needed.

Moreover, we did not simulate haze, clouds, and condensation processes, which can have a large impact on the abundances of certain elements and chemical species. Missing condensation would alter the upper atmosphere \ce{H2O} abundance in the warm 100x Solar and all cold cases. However, we expect minor depletions, because condensation does not substantially change the gaseous abundances in Jupiter \citep{Fegley1994Icar..110..117F}. Even so, we expect indirect effects of condensation via radiatively active clouds.
Radiatively active clouds \citep{Parmentier2021MNRAS,Roman2021,Komacek2022PatchyClouds, Malsky2024ApJ} change the temperature structure and atmospheric dynamics additionally. Especially, WD-1856b has higher \ce{H2O} abundances in both 100x Solar cases which affects several wavelength ranges. Firstly, radiatively active high-altitude clouds on the dayside can scatter at short wavelengths and increase the albedo, which lowers the amount of absorbed incoming radiation \citep{parmentier2016}. Secondly, radiatively active clouds scatter and absorb in the thermal wavelengths, which can cause a greenhouse effect \citep[e.g.][]{Mohandas_2018}. The simulation of clouds can help explain the variability in observations \citep[e.g. direct imaging of brown dwarfs][]{Tan2019Showman}.

\section{Summary and Conclusions}
\label{sec:conc_WD}

Insights into the Earth's future can be gained from White Dwarf (WD) systems that traverse the red giant branch. The first and only detected transiting planetary mass object in such a system \citep{Vanderburg2020Natur.585..363V} is the cold Jupiter WD-1856b+534b (WD-1856b). Recent JWST observation have targeted WD-1856b \citep{MacDonald2021jwst.prop.2358M, limbach2025thermalemissionconfirmationfrigid}, with more JWST observations expected in the future.

In order to support the interpretation of these observational data, we performed simulations of WD-1856b using a 3D general circulation model (GCM) and post-processed the results to produce synthetic emission spectra. We used the Exo-FMS GCM with a correlated-k radiative transfer (RT) scheme \citep{amundsen2016,Amundsen2017} and mixing length theory (MLT). Additionally, we computed the chemical abundances of 13 chemical species inside the GCM with the miniature chemical kinetics model mini-chem \citep{Tsai2022, Lee2023minichemicalscheme}. Because there are uncertainties in the metallicity and internal temperature of WD-1856b, we run simulations with 1x, 10x, and 100x Solar compositions, and with $T_{\rm int}$ = 100 and 500 K. The chemical abundances were used to compute the non-constant specific heat capacity that was used for the MLT and for the heat rates. In the post-processing, we generated synthetic emission and phase curves using the gCMCRT model \citep{Lee2022ApJ...929..180L_gCMCRT} and additional chemical equilibrium abundances with FastChem 2 \citep{Stock2022MNRAS.517.4070S} based on the GCM outputs. Regarding the temperature structure, we compared our T-p profiles with the 1D atmosphere models of ATMO2020 by \citet{Phillips2020A&A...637A..38P} and of Sonora Bobcat by \citet{Marley2021ApJ...920...85M}. Because WD-1856b is a cold gas giant similar to Jupiter, we compared both planets. Our cold 1x solar case shows similar temperatures at $p=$10$^{8}$~Pa as Jupiter. At lower pressures, the higher incident radiation leads to higher temperatures in WD-1856b.

Our results show the temperature structure, dynamics, chemical abundances and spectra of WD-1856b at 1x, 10x and 100x Solar composition, and at $T_{\rm int}$ = 100 and 500 K, respectively. We predicted a temperature structure with higher lapse rates than the 1D models of ATMO2020 and Sonora Bobcat did. The differences in the prediction arise from incoming radiation, non-constant heat capacity in the MLT (adiabatic lapse rates), and in the heating rates, from the 3D advection, and from the more accurate gravity. The higher Solar metallicities lead to higher temperatures in the deep atmosphere and lower temperatures in the upper atmosphere. The lower temperature and higher \ce{H2O} abundances shrink the "spread" (difference of dew point and the local temperature) so that \ce{H2O} cloud formation is possible in the 100x Solar case at $T_{\rm int}$ = 500 K and in all cases with $T_{\rm int}$ = 100 K. Hence, we expect that \ce{H2O} is slightly reduced in the cold trap in the upper atmosphere. The decreasing order of chemical species are line with \citet{Rensen2023RemS...15..841R}. The abundances in the 10x Solar case agree better with their study than in the 1x Solar case.

At $T_{\rm int}$ = 500 K, higher metallicities increase the wind speeds and the vertical extent of the jets in WD-1856b. However, the zonal winds remain below 200 m s$^{-1}$ at pressures $p$ < 100 Pa. In the 100x solar case, the overturning circulation cycles at higher rates than in the 10x Solar case, which increases the advective mixing. But we do not see this trend in an increasing circulation strength at lower metallicities.

At $T_{\rm int}$ = 100 K, the winds and the overturning circulation strengthen overall compared to the hotter case. The zonal winds reach speeds above 200 m s$^{-1}$ in all cold cases (even above 600 m s$^{-1}$ at 10x Solar). The 1x Solar case evolves the strongest winds and overturning circulation. The weak temperature gradient (WTG) parameter decreases significantly in the 100x solar case. However, there is no trend in decreasing WTG parameters or in increasing circulation strengths at lower metallicities. Perhaps longer simulations can evolve more distinct or even totally new circulation systems. For instance, \citet{Sainsbury2023MNRAS.524.1316S} experienced a substantial change in the jet systems in a simulation run of up to 50'000 days. Moreover, the shallow-water theory is not suitable for the large depth of this planet. Nonetheless, $\Lambda$ may describe the dynamical behaviour of the primitive equation set in the dynamical core of Exo-FMS GCM when applied to WD-1856b.

Regarding chemistry, mini-chem and FastChem 2 agree well on the abundances of 13 chemical species in the deeper atmosphere. In the upper atmosphere, the abundances produced by mini-chem do not change vertically above quench pressure, whereas abundances produced by the chemical equilibrium model FastChem 2 are strongly dependent on the pressure and temperature.

At $T_{\rm int}$ = 500 K, the most abundant chemical species apart from He and \ce{H2} are \ce{H2O}, CO and \ce{CH4}.
At Solar composition, the abundances of these elements are in the range of 100 and to 1'000 ppm. At 100x Solar composition, \ce{H2O} and CO increase to a few percentages, whereas \ce{CH4} remains closely below 1'000 ppm in the upper atmosphere. \ce{H2O} might be removed by cloud formation in the 100x Solar case to some degree.

At $T_{\rm int}$ = 100 K, \ce{H2O} and \ce{CH4} are the most abundant species apart from \ce{H2} and He. They increase from around 400 ppm to several percentages when we increase the metallicity by 2 magnitudes. In the Solar case, \ce{N2} and \ce{NH3} are the next fewer chemical species. At 100x Solar, \ce{N2} and CO may reach abundances up to 5'000 ppm. \ce{H2O} is likely removed by cloud formation and rain out.

Regarding spectral fluxes, we expect that absorptions in the bands of \ce{CH4} and \ce{H2O} are more pronounced in the colder cases, especially at 100x Solar. However, the absorption characteristics of \ce{H2O} is probably more reduced than we predicted. If WD-1856b has a higher $T_{\rm int}$, then we expect more absorption characteristics of CO and \ce{CO2}.

Therefore, the abundance ratios of \ce{CH4}, \ce{H2O} to CO and \ce{CO2} are good indicators of $T_{\rm int}$ in WD-1856b. A good indicator of the metallicity in WD-1856b is the abundance ratio of \ce{H2O} to CO at high $T_{\rm int}$. Whereas at low $T_{\rm int}$, such an indicator is the abundance ratio of \ce{N2} to \ce{NH3}.

Moreover, the temperature structures of our colder cases are in line with \citet{limbach2025thermalemissionconfirmationfrigid}. We assume that a higher $T_{\rm int}$ could evolve feasible T-p profiles to match their estimated effective temperature of 184 K as well. It could be worth running simulations with more sophisticated models and with $T_{\rm int}$ $\sim$ 200 K.

Future GCM investigations may extend this study by adding a microphysical cloud scheme such as CARMA \citep{Gao2018,Powell2018} and by adding radiative feedback from clouds \citep{Roman2021,Komacek2022PatchyClouds, Malsky2024ApJ}. Additionally, GCMs may simulated several more thousands of days than we have here. Another change may come from interactive photochemistry or photoionisation by the WD. 

Future observations targeting WD-1856b can specify our finding in more detail. They can help better understand how the RGB phase affects the $T_{\rm int}$ and the metallicity of planets in the habitable zone around a WD.

\section*{Acknowledgements}
P. Noti and E.K.H. Lee are supported by the SNSF Ambizione Fellowship grant (\#193448), a scheme of the Swiss National Science Foundation. Calculations were performed on UBELIX (\url{http://www.id.unibe.ch/hpc}), the HPC cluster at the Universit\"at Bern. We thank the IT Service Office (Ubelix cluster), the Physikalisches Institut and the Center for Space and Habitability at the Universit\"at of Bern for their services. The HPC support staff at the Universit\"at Bern is highly acknowledged. Data and plots were processed and produced using PYTHON version 3.9 \citep{van1995python} and the community open-source PYTHON packages \emph{Bokeh} \citep{bokeh}, \emph{Matplotlib} \citep{hunter2007}, \emph{cartopy} \citep{cartopy1,cartopy2}, \emph{jupyter} \citep{kluyver2016}, \emph{NumPy} \citep{harris2020}, \emph{pandas} \citep{reback2020pandas}, \emph{SciPy} \citep{jones2001}, \emph{seaborn} \citep{waskom2021}, \emph{windspharm} \citep{dawson2016} and \emph{xarray} \citet{hoyer2017}. Parts of this work have been carried out within the framework of the NCCR PlanetS supported by the Swiss National Science Foundation under grants 51NF40\_182901 and 51NF40\_205606.

The HPC support staff at the Universit\"at Bern is highly acknowledged.
Calculations were performed on UBELIX (\url{http://www.id.unibe.ch/hpc}), the HPC cluster at the Universit\"at Bern.

\section*{Data Availability}
The 1D radiative-transfer, mini-chem, gCMCRT and various other source codes are available on GitHub: \url{https://github.com/ELeeAstro}.
The GGM simulation output data (NetCDF format), gCMCRT and FastChem 2 output data are available on Zenodo under
\href{https://zenodo.org/records/15357129?token=eyJhbGciOiJIUzUxMiJ9.eyJpZCI6IjA2OTIzMGMzLTYzMDctNDVkOC05ZGVmLTM5YWU5MWYyM2MwNSIsImRhdGEiOnt9LCJyYW5kb20iOiJiODFlZTFkNjk2MjllYjllOGM3NzYzYzM3MTVlZjUxMSJ9.paJ53nmALzl4yOMAA84w6wtwtZqJGaw5v-QeEGO0UrMLGv_huWFDOh9iJIDAboOel7pcW-EpjeZ33FW7UDzYCQ}{https://doi.org/10.5281/zenodo.15357129}.
All other data is available upon request to the lead author.


\bibliographystyle{mnras}
\bibliography{main} 

\begin{thebibliography}{}
\makeatletter
\relax
\def\mn@urlcharsother{\let\do\@makeother \do\$\do\&\do\#\do\^\do\_\do\%\do\~}
\def\mn@doi{\begingroup\mn@urlcharsother \@ifnextchar [ {\mn@doi@} {\mn@doi@[]}}
\def\mn@doi@[#1]#2{\def\@tempa{#1}\ifx\@tempa\@empty \href {http://dx.doi.org/#2} {doi:#2}\else \href {http://dx.doi.org/#2} {#1}\fi \endgroup}
\def\mn@eprint#1#2{\mn@eprint@#1:#2::\@nil}
\def\mn@eprint@arXiv#1{\href {http://arxiv.org/abs/#1} {{\tt arXiv:#1}}}
\def\mn@eprint@dblp#1{\href {http://dblp.uni-trier.de/rec/bibtex/#1.xml} {dblp:#1}}
\def\mn@eprint@#1:#2:#3:#4\@nil{\def\@tempa {#1}\def\@tempb {#2}\def\@tempc {#3}\ifx \@tempc \@empty \let \@tempc \@tempb \let \@tempb \@tempa \fi \ifx \@tempb \@empty \def\@tempb {arXiv}\fi \@ifundefined {mn@eprint@\@tempb}{\@tempb:\@tempc}{\expandafter \expandafter \csname mn@eprint@\@tempb\endcsname \expandafter{\@tempc}}}

\bibitem[\protect\citeauthoryear{{Agol}}{{Agol}}{2011}]{Agol2011ApJ...731L..31A}
{Agol} E.,  2011, \mn@doi [\apjl] {10.1088/2041-8205/731/2/L31}, \href {https://ui.adsabs.harvard.edu/abs/2011ApJ...731L..31A} {731, L31}

\bibitem[\protect\citeauthoryear{{Amundsen} et~al.,}{{Amundsen} et~al.}{2016}]{amundsen2016}
{Amundsen} D.~S.,  et~al., 2016, \mn@doi [\aap] {10.1051/0004-6361/201629183}, \href {https://ui.adsabs.harvard.edu/abs/2016A&A...595A..36A} {595, A36}

\bibitem[\protect\citeauthoryear{{Amundsen}, {Tremblin}, {Manners}, {Baraffe}  \& {Mayne}}{{Amundsen} et~al.}{2017}]{Amundsen2017}
{Amundsen} D.~S.,  {Tremblin} P.,  {Manners} J.,  {Baraffe} I.,   {Mayne} N.~J.,  2017, \mn@doi [\aap] {10.1051/0004-6361/201629322}, \href {https://ui.adsabs.harvard.edu/abs/2017A&A...598A..97A} {598, A97}

\bibitem[\protect\citeauthoryear{{Asplund}, {Amarsi}  \& {Grevesse}}{{Asplund} et~al.}{2021}]{Asplund2021A&A...653A.141A}
{Asplund} M.,  {Amarsi} A.~M.,   {Grevesse} N.,  2021, \mn@doi [\aap] {10.1051/0004-6361/202140445}, \href {https://ui.adsabs.harvard.edu/abs/2021A&A...653A.141A} {653, A141}

\bibitem[\protect\citeauthoryear{{Barnes} \& {Heller}}{{Barnes} \& {Heller}}{2013}]{Barnes2013AsBio..13..279B}
{Barnes} R.,  {Heller} R.,  2013, \mn@doi [Astrobiology] {10.1089/ast.2012.0867}, \href {https://ui.adsabs.harvard.edu/abs/2013AsBio..13..279B} {13, 279}

\bibitem[\protect\citeauthoryear{{Barshay} \& {Lewis}}{{Barshay} \& {Lewis}}{1978}]{Barshay1978Icar...33..593B}
{Barshay} S.~S.,  {Lewis} J.~S.,  1978, \mn@doi [\icarus] {10.1016/0019-1035(78)90192-6}, \href {https://ui.adsabs.harvard.edu/abs/1978Icar...33..593B} {33, 593}

\bibitem[\protect\citeauthoryear{{Becker}, {Seligman}, {Adams}  \& {Styczinski}}{{Becker} et~al.}{2023}]{Becker2023ApJ...945L..24B}
{Becker} J.,  {Seligman} D.~Z.,  {Adams} F.~C.,   {Styczinski} M.~J.,  2023, \mn@doi [\apjl] {10.3847/2041-8213/acbe44}, \href {https://ui.adsabs.harvard.edu/abs/2023ApJ...945L..24B} {945, L24}

\bibitem[\protect\citeauthoryear{{Bokeh Development Team}}{{Bokeh Development Team}}{2022}]{bokeh}
{Bokeh Development Team} 2022, Bokeh: Python library for interactive visualization.
\url {https://bokeh.org/}

\bibitem[\protect\citeauthoryear{{Bolton} et~al.,}{{Bolton} et~al.}{2017}]{Bolton2017Sci...356..821B}
{Bolton} S.~J.,  et~al., 2017, \mn@doi [Science] {10.1126/science.aal2108}, \href {https://ui.adsabs.harvard.edu/abs/2017Sci...356..821B} {356, 821}

\bibitem[\protect\citeauthoryear{{Bosman}, {Cridland}  \& {Miguel}}{{Bosman} et~al.}{2019}]{Bosman2019A&A...632L..11B}
{Bosman} A.~D.,  {Cridland} A.~J.,   {Miguel} Y.,  2019, \mn@doi [\aap] {10.1051/0004-6361/201936827}, \href {https://ui.adsabs.harvard.edu/abs/2019A&A...632L..11B} {632, L11}

\bibitem[\protect\citeauthoryear{{Burrows} \& {Liebert}}{{Burrows} \& {Liebert}}{1993}]{Burrows1993RvMP...65..301B}
{Burrows} A.,  {Liebert} J.,  1993, \mn@doi [Reviews of Modern Physics] {10.1103/RevModPhys.65.301}, \href {https://ui.adsabs.harvard.edu/abs/1993RvMP...65..301B} {65, 301}

\bibitem[\protect\citeauthoryear{{Carlson}, {Prather}  \& {Rossow}}{{Carlson} et~al.}{1987}]{Carlson1987ApJ...322..559C}
{Carlson} B.~E.,  {Prather} M.~J.,   {Rossow} W.~B.,  1987, \mn@doi [\apj] {10.1086/165751}, \href {https://ui.adsabs.harvard.edu/abs/1987ApJ...322..559C} {322, 559}

\bibitem[\protect\citeauthoryear{{Carri{\'o}n-Gonz{\'a}lez} et~al.,}{{Carri{\'o}n-Gonz{\'a}lez} et~al.}{2023}]{Carrion2023A&A...674L...3C}
{Carri{\'o}n-Gonz{\'a}lez} {\'O}.,  et~al., 2023, \mn@doi [\aap] {10.1051/0004-6361/202346621}, \href {https://ui.adsabs.harvard.edu/abs/2023A&A...674L...3C} {674, L3}

\bibitem[\protect\citeauthoryear{{Casewell} et~al.,}{{Casewell} et~al.}{2015}]{Casewell2015}
{Casewell} S.~L.,  et~al., 2015, \mn@doi [\mnras] {10.1093/mnras/stu2721}, \href {https://ui.adsabs.harvard.edu/abs/2015MNRAS.447.3218C} {447, 3218}

\bibitem[\protect\citeauthoryear{{Chamandy}, {Blackman}, {Nordhaus}  \& {Wilson}}{{Chamandy} et~al.}{2021}]{Chamandy2021MNRAS.502L.110C}
{Chamandy} L.,  {Blackman} E.~G.,  {Nordhaus} J.,   {Wilson} E.,  2021, \mn@doi [\mnras] {10.1093/mnrasl/slab017}, \href {https://ui.adsabs.harvard.edu/abs/2021MNRAS.502L.110C} {502, L110}

\bibitem[\protect\citeauthoryear{{Charnay}, {Blain}, {B{\'e}zard}, {Leconte}, {Turbet}  \& {Falco}}{{Charnay} et~al.}{2021}]{Charnay2021A&A...646A.171C}
{Charnay} B.,  {Blain} D.,  {B{\'e}zard} B.,  {Leconte} J.,  {Turbet} M.,   {Falco} A.,  2021, \mn@doi [\aap] {10.1051/0004-6361/202039525}, \href {https://ui.adsabs.harvard.edu/abs/2021A&A...646A.171C} {646, A171}

\bibitem[\protect\citeauthoryear{{Charney}}{{Charney}}{1963}]{Charney1963JAtS...20..607C}
{Charney} J.~G.,  1963, \mn@doi [Journal of the Atmospheric Sciences] {10.1175/1520-0469(1963)020<0607:ANOLSM>2.0.CO;2}, \href {https://ui.adsabs.harvard.edu/abs/1963JAtS...20..607C} {20, 607}

\bibitem[\protect\citeauthoryear{{Chase}}{{Chase}}{1986}]{Chase1986jtt..book.....C}
{Chase} M.~W.,  1986, {JANAF thermochemical tables}

\bibitem[\protect\citeauthoryear{{Christie} et~al.,}{{Christie} et~al.}{2022}]{christie2022}
{Christie} D.~A.,  et~al., 2022, \mn@doi [\psj] {10.3847/PSJ/ac9dfe}, \href {https://ui.adsabs.harvard.edu/abs/2022PSJ.....3..261C} {3, 261}

\bibitem[\protect\citeauthoryear{{Chubb}, {Tennyson}  \& {Yurchenko}}{{Chubb} et~al.}{2020}]{Chubb2020}
{Chubb} K.~L.,  {Tennyson} J.,   {Yurchenko} S.~N.,  2020, \mn@doi [\mnras] {10.1093/mnras/staa229}, \href {https://ui.adsabs.harvard.edu/abs/2020MNRAS.493.1531C} {493, 1531}

\bibitem[\protect\citeauthoryear{{Coles}, {Yurchenko}  \& {Tennyson}}{{Coles} et~al.}{2019}]{Coles2019}
{Coles} P.~A.,  {Yurchenko} S.~N.,   {Tennyson} J.,  2019, \mn@doi [\mnras] {10.1093/mnras/stz2778}, \href {https://ui.adsabs.harvard.edu/abs/2019MNRAS.490.4638C} {490, 4638}

\bibitem[\protect\citeauthoryear{Dawson}{Dawson}{2016}]{dawson2016}
Dawson A.,  2016, \mn@doi [Journal of Open Research Software] {10.5334/jors.129}

\bibitem[\protect\citeauthoryear{{Debes} \& {Sigurdsson}}{{Debes} \& {Sigurdsson}}{2002}]{Debes2002ApJ...572..556D}
{Debes} J.~H.,  {Sigurdsson} S.,  2002, \mn@doi [\apj] {10.1086/340291}, \href {https://ui.adsabs.harvard.edu/abs/2002ApJ...572..556D} {572, 556}

\bibitem[\protect\citeauthoryear{{Deitrick}, {Mendon{\c{c}}a}, {Schroffenegger}, {Grimm}, {Tsai}  \& {Heng}}{{Deitrick} et~al.}{2020}]{deitrick2020}
{Deitrick} R.,  {Mendon{\c{c}}a} J.~M.,  {Schroffenegger} U.,  {Grimm} S.~L.,  {Tsai} S.-M.,   {Heng} K.,  2020, \mn@doi [\apjs] {10.3847/1538-4365/ab930e}, \href {https://ui.adsabs.harvard.edu/abs/2020ApJS..248...30D} {248, 30}

\bibitem[\protect\citeauthoryear{{Duer} et~al.,}{{Duer} et~al.}{2021}]{Duer2021GeoRL..4895651D}
{Duer} K.,  et~al., 2021, \mn@doi [\grl] {10.1029/2021GL095651}, \href {https://ui.adsabs.harvard.edu/abs/2021GeoRL..4895651D} {48, e95651}

\bibitem[\protect\citeauthoryear{Elson et~al.,}{Elson et~al.}{2022}]{cartopy2}
Elson P.,  et~al., 2022, SciTools/cartopy: v0.21.1, \mn@doi{10.5281/zenodo.7430317}, \url {https://doi.org/10.5281/zenodo.7430317}

\bibitem[\protect\citeauthoryear{{Fegley} \& {Lodders}}{{Fegley} \& {Lodders}}{1994}]{Fegley1994Icar..110..117F}
{Fegley} Jr. B.,  {Lodders} K.,  1994, \mn@doi [\icarus] {10.1006/icar.1994.1111}, \href {https://ui.adsabs.harvard.edu/abs/1994Icar..110..117F} {110, 117}

\bibitem[\protect\citeauthoryear{{Fletcher} et~al.,}{{Fletcher} et~al.}{2021}]{Fletcher2021JGRE..12606858F}
{Fletcher} L.~N.,  et~al., 2021, \mn@doi [Journal of Geophysical Research (Planets)] {10.1029/2021JE006858}, \href {https://ui.adsabs.harvard.edu/abs/2021JGRE..12606858F} {126, e06858}

\bibitem[\protect\citeauthoryear{{Fossati}, {Bagnulo}, {Haswell}, {Patel}, {Busuttil}, {Kowalski}, {Shulyak}  \& {Sterzik}}{{Fossati} et~al.}{2012}]{Fossati2012ApJ...757L..15F}
{Fossati} L.,  {Bagnulo} S.,  {Haswell} C.~A.,  {Patel} M.~R.,  {Busuttil} R.,  {Kowalski} P.~M.,  {Shulyak} D.~V.,   {Sterzik} M.~F.,  2012, \mn@doi [\apjl] {10.1088/2041-8205/757/1/L15}, \href {https://ui.adsabs.harvard.edu/abs/2012ApJ...757L..15F} {757, L15}

\bibitem[\protect\citeauthoryear{{Gao}, {Marley}  \& {Ackerman}}{{Gao} et~al.}{2018}]{Gao2018}
{Gao} P.,  {Marley} M.~S.,   {Ackerman} A.~S.,  2018, \mn@doi [\apj] {10.3847/1538-4357/aab0a1}, \href {https://ui.adsabs.harvard.edu/abs/2018ApJ...855...86G} {855, 86}

\bibitem[\protect\citeauthoryear{{Garc{\'\i}a-Melendo}, {P{\'e}rez-Hoyos}, {S{\'a}nchez-Lavega}  \& {Hueso}}{{Garc{\'\i}a-Melendo} et~al.}{2011}]{Garcia2011Icar..215...62G}
{Garc{\'\i}a-Melendo} E.,  {P{\'e}rez-Hoyos} S.,  {S{\'a}nchez-Lavega} A.,   {Hueso} R.,  2011, \mn@doi [\icarus] {10.1016/j.icarus.2011.07.005}, \href {https://ui.adsabs.harvard.edu/abs/2011Icar..215...62G} {215, 62}

\bibitem[\protect\citeauthoryear{{Grassi} et~al.,}{{Grassi} et~al.}{2020}]{Grassi2020JGRE..12506206G}
{Grassi} D.,  et~al., 2020, \mn@doi [Journal of Geophysical Research (Planets)] {10.1029/2019JE006206}, \href {https://ui.adsabs.harvard.edu/abs/2020JGRE..12506206G} {125, e06206}

\bibitem[\protect\citeauthoryear{{Guillot} \& {Morel}}{{Guillot} \& {Morel}}{1995}]{Guillot1995A&AS..109..109G}
{Guillot} T.,  {Morel} P.,  1995, \aaps, \href {https://ui.adsabs.harvard.edu/abs/1995A&AS..109..109G} {109, 109}

\bibitem[\protect\citeauthoryear{{Guillot}, {Burrows}, {Hubbard}, {Lunine}  \& {Saumon}}{{Guillot} et~al.}{1996}]{Guillot1996ApJ...459L..35G}
{Guillot} T.,  {Burrows} A.,  {Hubbard} W.~B.,  {Lunine} J.~I.,   {Saumon} D.,  1996, \mn@doi [\apjl] {10.1086/309935}, \href {https://ui.adsabs.harvard.edu/abs/1996ApJ...459L..35G} {459, L35}

\bibitem[\protect\citeauthoryear{{Guillot}, {Stevenson}, {Atreya}, {Bolton}  \& {Becker}}{{Guillot} et~al.}{2020a}]{Guillot2020JGRE..12506403G}
{Guillot} T.,  {Stevenson} D.~J.,  {Atreya} S.~K.,  {Bolton} S.~J.,   {Becker} H.~N.,  2020a, \mn@doi [Journal of Geophysical Research (Planets)] {10.1029/2020JE00640310.1002/essoar.10502154.1}, \href {https://ui.adsabs.harvard.edu/abs/2020JGRE..12506403G} {125, e06403}

\bibitem[\protect\citeauthoryear{{Guillot} et~al.,}{{Guillot} et~al.}{2020b}]{Guillot2020JGRE..12506404G}
{Guillot} T.,  et~al., 2020b, \mn@doi [Journal of Geophysical Research (Planets)] {10.1029/2020JE00640410.1002/essoar.10502179.1}, \href {https://ui.adsabs.harvard.edu/abs/2020JGRE..12506404G} {125, e06404}

\bibitem[\protect\citeauthoryear{{Guillot}, {Fletcher}, {Helled}, {Ikoma}, {Line}  \& {Paramentier}}{{Guillot} et~al.}{2023}]{Guillot2023ASPC..534..947G}
{Guillot} T.,  {Fletcher} L.~N.,  {Helled} R.,  {Ikoma} M.,  {Line} M.~R.,   {Paramentier} V.,  2023, in {Inutsuka} S.,  {Aikawa} Y.,  {Muto} T.,  {Tomida} K.,   {Tamura} M.,  eds,  Astronomical Society of the Pacific Conference Series Vol. 534, Protostars and Planets VII. p.~947 (\mn@eprint {arXiv} {2205.04100}), \mn@doi{10.48550/arXiv.2205.04100}

\bibitem[\protect\citeauthoryear{{Hargreaves}, {Gordon}, {Kochanov}  \& {Rothman}}{{Hargreaves} et~al.}{2019}]{Hargreaves2019}
{Hargreaves} R.,  {Gordon} I.,  {Kochanov} R.,   {Rothman} L.,  2019, in EPSC-DPS Joint Meeting 2019. pp EPSC--DPS2019--919

\bibitem[\protect\citeauthoryear{{Hargreaves}, {Gordon}, {Rey}, {Nikitin}, {Tyuterev}, {Kochanov}  \& {Rothman}}{{Hargreaves} et~al.}{2020}]{Hargreaves2020}
{Hargreaves} R.~J.,  {Gordon} I.~E.,  {Rey} M.,  {Nikitin} A.~V.,  {Tyuterev} V.~G.,  {Kochanov} R.~V.,   {Rothman} L.~S.,  2020, \mn@doi [\apjs] {10.3847/1538-4365/ab7a1a}, \href {https://ui.adsabs.harvard.edu/abs/2020ApJS..247...55H} {247, 55}

\bibitem[\protect\citeauthoryear{{Harris}, {Tennyson}, {Kaminsky}, {Pavlenko}  \& {Jones}}{{Harris} et~al.}{2006}]{Harris2006}
{Harris} G.~J.,  {Tennyson} J.,  {Kaminsky} B.~M.,  {Pavlenko} Y.~V.,   {Jones} H.~R.~A.,  2006, \mn@doi [\mnras] {10.1111/j.1365-2966.2005.09960.x}, \href {https://ui.adsabs.harvard.edu/abs/2006MNRAS.367..400H} {367, 400}

\bibitem[\protect\citeauthoryear{{Harris} et~al.,}{{Harris} et~al.}{2020}]{harris2020}
{Harris} C.~R.,  et~al., 2020, \mn@doi [\nat] {10.1038/s41586-020-2649-2}, \href {https://ui.adsabs.harvard.edu/abs/2020Natur.585..357H} {585, 357}

\bibitem[\protect\citeauthoryear{{Helled} \& {Lunine}}{{Helled} \& {Lunine}}{2014}]{Helled2014MNRAS.441.2273H}
{Helled} R.,  {Lunine} J.,  2014, \mn@doi [\mnras] {10.1093/mnras/stu516}, \href {https://ui.adsabs.harvard.edu/abs/2014MNRAS.441.2273H} {441, 2273}

\bibitem[\protect\citeauthoryear{{Howard} \& {Guillot}}{{Howard} \& {Guillot}}{2023}]{Howard2023A&A...672L...1H}
{Howard} S.,  {Guillot} T.,  2023, \mn@doi [\aap] {10.1051/0004-6361/202244851}, \href {https://ui.adsabs.harvard.edu/abs/2023A&A...672L...1H} {672, L1}

\bibitem[\protect\citeauthoryear{{Howard} et~al.,}{{Howard} et~al.}{2023}]{Howard2023A&A...672A..33H}
{Howard} S.,  et~al., 2023, \mn@doi [\aap] {10.1051/0004-6361/202245625}, \href {https://ui.adsabs.harvard.edu/abs/2023A&A...672A..33H} {672, A33}

\bibitem[\protect\citeauthoryear{Hoyer \& Hamman}{Hoyer \& Hamman}{2017}]{hoyer2017}
Hoyer S.,  Hamman J.,  2017, \mn@doi [Journal of Open Research Software] {10.5334/jors.148}

\bibitem[\protect\citeauthoryear{{Hunter}}{{Hunter}}{2007}]{hunter2007}
{Hunter} J.~D.,  2007, \mn@doi [Computing in Science and Engineering] {10.1109/MCSE.2007.55}, \href {https://ui.adsabs.harvard.edu/abs/2007CSE.....9...90H} {9, 90}

\bibitem[\protect\citeauthoryear{{Ingersoll} et~al.,}{{Ingersoll} et~al.}{2017}]{Ingersoll2017GeoRL..44.7676I}
{Ingersoll} A.~P.,  et~al., 2017, \mn@doi [\grl] {10.1002/2017GL074277}, \href {https://ui.adsabs.harvard.edu/abs/2017GeoRL..44.7676I} {44, 7676}

\bibitem[\protect\citeauthoryear{Jones, Oliphant, Peterson  et~al.}{Jones et~al.}{2001}]{jones2001}
Jones E.,  Oliphant T.,  Peterson P.,   et~al., 2001, SciPy: Open source scientific tools for Python, \url {https://www.scipy.org}

\bibitem[\protect\citeauthoryear{{Karkoschka}}{{Karkoschka}}{2011}]{Karkoschka2011Icar..215..439K}
{Karkoschka} E.,  2011, \mn@doi [\icarus] {10.1016/j.icarus.2011.05.013}, \href {https://ui.adsabs.harvard.edu/abs/2011Icar..215..439K} {215, 439}

\bibitem[\protect\citeauthoryear{{Karman} et~al.,}{{Karman} et~al.}{2019}]{Karman2019}
{Karman} T.,  et~al., 2019, \mn@doi [\icarus] {10.1016/j.icarus.2019.02.034}, \href {https://ui.adsabs.harvard.edu/abs/2019Icar..328..160K} {328, 160}

\bibitem[\protect\citeauthoryear{{Kataria}, {Showman}, {Lewis}, {Fortney}, {Marley}  \& {Freedman}}{{Kataria} et~al.}{2013}]{Kataria2013}
{Kataria} T.,  {Showman} A.~P.,  {Lewis} N.~K.,  {Fortney} J.~J.,  {Marley} M.~S.,   {Freedman} R.~S.,  2013, \mn@doi [\apj] {10.1088/0004-637X/767/1/76}, \href {https://ui.adsabs.harvard.edu/abs/2013ApJ...767...76K} {767, 76}

\bibitem[\protect\citeauthoryear{{Kempton} et~al.,}{{Kempton} et~al.}{2023}]{Kempton2023Natur.620...67K}
{Kempton} E. M.~R.,  et~al., 2023, \mn@doi [\nat] {10.1038/s41586-023-06159-5}, \href {https://ui.adsabs.harvard.edu/abs/2023Natur.620...67K} {620, 67}

\bibitem[\protect\citeauthoryear{{Kluyver} et~al.,}{{Kluyver} et~al.}{2016}]{kluyver2016}
{Kluyver} T.,  et~al., 2016, in , IOS Press.
IOS Press Ebooks, pp 87--90, \mn@doi{10.3233/978-1-61499-649-1-87}

\bibitem[\protect\citeauthoryear{{Komacek}, {Tan}, {Gao}  \& {Lee}}{{Komacek} et~al.}{2022}]{Komacek2022PatchyClouds}
{Komacek} T.~D.,  {Tan} X.,  {Gao} P.,   {Lee} E. K.~H.,  2022, \mn@doi [\apj] {10.3847/1538-4357/ac7723}, \href {https://ui.adsabs.harvard.edu/abs/2022ApJ...934...79K} {934, 79}

\bibitem[\protect\citeauthoryear{{Kozakis}, {Lin}  \& {Kaltenegger}}{{Kozakis} et~al.}{2020}]{Kozakis2020ApJ...894L...6K}
{Kozakis} T.,  {Lin} Z.,   {Kaltenegger} L.,  2020, \mn@doi [\apjl] {10.3847/2041-8213/ab6f6a}, \href {https://ui.adsabs.harvard.edu/abs/2020ApJ...894L...6K} {894, L6}

\bibitem[\protect\citeauthoryear{{Lagos}, {Schreiber}, {Zorotovic}, {G{\"a}nsicke}, {Ronco}  \& {Hamers}}{{Lagos} et~al.}{2021}]{Lagos2021MNRAS.501..676L}
{Lagos} F.,  {Schreiber} M.~R.,  {Zorotovic} M.,  {G{\"a}nsicke} B.~T.,  {Ronco} M.~P.,   {Hamers} A.~S.,  2021, \mn@doi [\mnras] {10.1093/mnras/staa3703}, \href {https://ui.adsabs.harvard.edu/abs/2021MNRAS.501..676L} {501, 676}

\bibitem[\protect\citeauthoryear{{Lee}, {Parmentier}, {Hammond}, {Grimm}, {Kitzmann}, {Tan}, {Tsai}  \& {Pierrehumbert}}{{Lee} et~al.}{2021}]{lee2021}
{Lee} E. K.~H.,  {Parmentier} V.,  {Hammond} M.,  {Grimm} S.~L.,  {Kitzmann} D.,  {Tan} X.,  {Tsai} S.-M.,   {Pierrehumbert} R.~T.,  2021, \mn@doi [\mnras] {10.1093/mnras/stab1851}, \href {https://ui.adsabs.harvard.edu/abs/2021MNRAS.506.2695L} {506, 2695}

\bibitem[\protect\citeauthoryear{{Lee}, {Prinoth}, {Kitzmann}, {Tsai}, {Hoeijmakers}, {Borsato}  \& {Heng}}{{Lee} et~al.}{2022a}]{lee2022}
{Lee} E. K.~H.,  {Prinoth} B.,  {Kitzmann} D.,  {Tsai} S.-M.,  {Hoeijmakers} J.,  {Borsato} N.~W.,   {Heng} K.,  2022a, \mn@doi [\mnras] {10.1093/mnras/stac2246}, \href {https://ui.adsabs.harvard.edu/abs/2022MNRAS.517..240L} {517, 240}

\bibitem[\protect\citeauthoryear{{Lee} et~al.,}{{Lee} et~al.}{2022b}]{Lee2022ApJ...929..180L_gCMCRT}
{Lee} E. K.~H.,  et~al., 2022b, \mn@doi [\apj] {10.3847/1538-4357/ac61d6}, \href {https://ui.adsabs.harvard.edu/abs/2022ApJ...929..180L} {929, 180}

\bibitem[\protect\citeauthoryear{{Lee}, {Tan}  \& {Tsai}}{{Lee} et~al.}{2023a}]{Lee2023MNRAS.523.4477L}
{Lee} E. K.~H.,  {Tan} X.,   {Tsai} S.-M.,  2023a, \mn@doi [\mnras] {10.1093/mnras/stad1715}, \href {https://ui.adsabs.harvard.edu/abs/2023MNRAS.523.4477L} {523, 4477}

\bibitem[\protect\citeauthoryear{{Lee}, {Tsai}, {Hammond}  \& {Tan}}{{Lee} et~al.}{2023b}]{Lee2023minichemicalscheme}
{Lee} E. K.~H.,  {Tsai} S.-M.,  {Hammond} M.,   {Tan} X.,  2023b, \mn@doi [\aap] {10.1051/0004-6361/202245473}, \href {https://ui.adsabs.harvard.edu/abs/2023A&A...672A.110L} {672, A110}

\bibitem[\protect\citeauthoryear{{Lee}, {Tan}  \& {Tsai}}{{Lee} et~al.}{2024}]{Lee2024MNRAS.529.2686L}
{Lee} E. K.~H.,  {Tan} X.,   {Tsai} S.-M.,  2024, \mn@doi [\mnras] {10.1093/mnras/stae537}, \href {https://ui.adsabs.harvard.edu/abs/2024MNRAS.529.2686L} {529, 2686}

\bibitem[\protect\citeauthoryear{{Li}, {Gordon}, {Rothman}, {Tan}, {Hu}, {Kassi}, {Campargue}  \& {Medvedev}}{{Li} et~al.}{2015}]{Li2015}
{Li} G.,  {Gordon} I.~E.,  {Rothman} L.~S.,  {Tan} Y.,  {Hu} S.-M.,  {Kassi} S.,  {Campargue} A.,   {Medvedev} E.~S.,  2015, \mn@doi [The Astrophysical Journal Supplement Series] {10.1088/0067-0049/216/1/15}, \href {https://ui.adsabs.harvard.edu/#abs/2015ApJS..216...15L} {216, 15}

\bibitem[\protect\citeauthoryear{{Li} et~al.,}{{Li} et~al.}{2017}]{Li2017GeoRL..44.5317L}
{Li} C.,  et~al., 2017, \mn@doi [\grl] {10.1002/2017GL073159}, \href {https://ui.adsabs.harvard.edu/abs/2017GeoRL..44.5317L} {44, 5317}

\bibitem[\protect\citeauthoryear{{Li} et~al.,}{{Li} et~al.}{2020}]{Li2020NatAs...4..609LJuno}
{Li} C.,  et~al., 2020, \mn@doi [Nature Astronomy] {10.1038/s41550-020-1009-3}, \href {https://ui.adsabs.harvard.edu/abs/2020NatAs...4..609L} {4, 609}

\bibitem[\protect\citeauthoryear{{Li} et~al.,}{{Li} et~al.}{2024}]{Li2024Icar..41416028L}
{Li} C.,  et~al., 2024, \mn@doi [\icarus] {10.1016/j.icarus.2024.116028}, \href {https://ui.adsabs.harvard.edu/abs/2024Icar..41416028L} {414, 116028}

\bibitem[\protect\citeauthoryear{Limbach et~al.,}{Limbach et~al.}{2025}]{limbach2025thermalemissionconfirmationfrigid}
Limbach M.~A.,  et~al., 2025, Thermal Emission and Confirmation of the Frigid White Dwarf Exoplanet WD 1856+534b (\mn@eprint {arXiv} {2504.16982}), \url {https://arxiv.org/abs/2504.16982}

\bibitem[\protect\citeauthoryear{{Loeb} \& {Maoz}}{{Loeb} \& {Maoz}}{2013}]{Loeb2013MNRAS.432L..11L}
{Loeb} A.,  {Maoz} D.,  2013, \mn@doi [\mnras] {10.1093/mnrasl/slt026}, \href {https://ui.adsabs.harvard.edu/abs/2013MNRAS.432L..11L} {432, L11}

\bibitem[\protect\citeauthoryear{{MacDonald} et~al.,}{{MacDonald} et~al.}{2021}]{MacDonald2021jwst.prop.2358M}
{MacDonald} R.~J.,  et~al., 2021, {Under the Light of a Dead Star: Revealing the Atmospheric Composition of a White Dwarf Planet}, JWST Proposal. Cycle 1, ID. \#2358

\bibitem[\protect\citeauthoryear{{Maldonado}, {Villaver}, {Mustill}, {Ch{\'a}vez}  \& {Bertone}}{{Maldonado} et~al.}{2021}]{Maldonado2021MNRAS.501L..43M}
{Maldonado} R.~F.,  {Villaver} E.,  {Mustill} A.~J.,  {Ch{\'a}vez} M.,   {Bertone} E.,  2021, \mn@doi [\mnras] {10.1093/mnrasl/slaa193}, \href {https://ui.adsabs.harvard.edu/abs/2021MNRAS.501L..43M} {501, L43}

\bibitem[\protect\citeauthoryear{{Malsky}, {Rauscher}, {Roman}, {Lee}, {Beltz}, {Savel}, {Kempton}  \& {Cinque}}{{Malsky} et~al.}{2024}]{Malsky2024ApJ}
{Malsky} I.,  {Rauscher} E.,  {Roman} M.~T.,  {Lee} E. K.~H.,  {Beltz} H.,  {Savel} A.,  {Kempton} E. M.~R.,   {Cinque} L.,  2024, \mn@doi [\apj] {10.3847/1538-4357/ad0b70}, \href {https://ui.adsabs.harvard.edu/abs/2024ApJ...961...66M} {961, 66}

\bibitem[\protect\citeauthoryear{{Marley} et~al.,}{{Marley} et~al.}{2021}]{Marley2021ApJ...920...85M}
{Marley} M.~S.,  et~al., 2021, \mn@doi [\apj] {10.3847/1538-4357/ac141d}, \href {https://ui.adsabs.harvard.edu/abs/2021ApJ...920...85M} {920, 85}

\bibitem[\protect\citeauthoryear{{Matthews} et~al.,}{{Matthews} et~al.}{2024}]{Matthews2024Natur.633..789M}
{Matthews} E.~C.,  et~al., 2024, \mn@doi [\nat] {10.1038/s41586-024-07837-8}, \href {https://ui.adsabs.harvard.edu/abs/2024Natur.633..789M} {633, 789}

\bibitem[\protect\citeauthoryear{{Mayne}, {Drummond}, {Debras}, {Jaupart}, {Manners}, {Boutle}, {Baraffe}  \& {Kohary}}{{Mayne} et~al.}{2019}]{mayne2019}
{Mayne} N.~J.,  {Drummond} B.,  {Debras} F.,  {Jaupart} E.,  {Manners} J.,  {Boutle} I.~A.,  {Baraffe} I.,   {Kohary} K.,  2019, \mn@doi [\apj] {10.3847/1538-4357/aaf6e9}, \href {https://ui.adsabs.harvard.edu/abs/2019ApJ...871...56M} {871, 56}

\bibitem[\protect\citeauthoryear{{Met Office}}{{Met Office}}{2015}]{cartopy1}
{Met Office} 2010 - 2015, Cartopy: a cartographic python library with a Matplotlib interface.
Exeter, Devon, \url {https://scitools.org.uk/cartopy}

\bibitem[\protect\citeauthoryear{{Miguel} et~al.,}{{Miguel} et~al.}{2022}]{Miguel2022A&A...662A..18M}
{Miguel} Y.,  et~al., 2022, \mn@doi [\aap] {10.1051/0004-6361/202243207}, \href {https://ui.adsabs.harvard.edu/abs/2022A&A...662A..18M} {662, A18}

\bibitem[\protect\citeauthoryear{{Mohandas}, {Pessah}  \& {Heng}}{{Mohandas} et~al.}{2018}]{Mohandas_2018}
{Mohandas} G.,  {Pessah} M.~E.,   {Heng} K.,  2018, \mn@doi [\apj] {10.3847/1538-4357/aab35d}, \href {https://ui.adsabs.harvard.edu/abs/2018ApJ...858....1M} {858, 1}

\bibitem[\protect\citeauthoryear{{Mousis}, {Marboeuf}, {Lunine}, {Alibert}, {Fletcher}, {Orton}, {Pauzat}  \& {Ellinger}}{{Mousis} et~al.}{2009}]{Mousis2009ApJ...696.1348M}
{Mousis} O.,  {Marboeuf} U.,  {Lunine} J.~I.,  {Alibert} Y.,  {Fletcher} L.~N.,  {Orton} G.~S.,  {Pauzat} F.,   {Ellinger} Y.,  2009, \mn@doi [\apj] {10.1088/0004-637X/696/2/1348}, \href {https://ui.adsabs.harvard.edu/abs/2009ApJ...696.1348M} {696, 1348}

\bibitem[\protect\citeauthoryear{{Mu{\~n}oz} \& {Petrovich}}{{Mu{\~n}oz} \& {Petrovich}}{2020}]{Munoz2020ApJ...904L...3M}
{Mu{\~n}oz} D.~J.,  {Petrovich} C.,  2020, \mn@doi [\apjl] {10.3847/2041-8213/abc564}, \href {https://ui.adsabs.harvard.edu/abs/2020ApJ...904L...3M} {904, L3}

\bibitem[\protect\citeauthoryear{{Noti}, {Lee}, {Deitrick}  \& {Hammond}}{{Noti} et~al.}{2023}]{Noti2023MNRAS.524.3396N}
{Noti} P.~A.,  {Lee} E. K.~H.,  {Deitrick} R.,   {Hammond} M.,  2023, \mn@doi [\mnras] {10.1093/mnras/stad2042}, \href {https://ui.adsabs.harvard.edu/abs/2023MNRAS.524.3396N} {524, 3396}

\bibitem[\protect\citeauthoryear{{O'Connor}, {Liu}  \& {Lai}}{{O'Connor} et~al.}{2021}]{OConnor2021MNRAS.501..507O}
{O'Connor} C.~E.,  {Liu} B.,   {Lai} D.,  2021, \mn@doi [\mnras] {10.1093/mnras/staa3723}, \href {https://ui.adsabs.harvard.edu/abs/2021MNRAS.501..507O} {501, 507}

\bibitem[\protect\citeauthoryear{{{\"O}berg} \& {Wordsworth}}{{{\"O}berg} \& {Wordsworth}}{2019}]{Oberg2019AJ....158..194O}
{{\"O}berg} K.~I.,  {Wordsworth} R.,  2019, \mn@doi [\aj] {10.3847/1538-3881/ab46a8}, \href {https://ui.adsabs.harvard.edu/abs/2019AJ....158..194O} {158, 194}

\bibitem[\protect\citeauthoryear{{Ohno} \& {Zhang}}{{Ohno} \& {Zhang}}{2019}]{Ohno2019ApJ...874....1O}
{Ohno} K.,  {Zhang} X.,  2019, \mn@doi [\apj] {10.3847/1538-4357/ab06cc}, \href {https://ui.adsabs.harvard.edu/abs/2019ApJ...874....1O} {874, 1}

\bibitem[\protect\citeauthoryear{{Parmentier}, {Fortney}, {Showman}, {Morley}  \& {Marley}}{{Parmentier} et~al.}{2016}]{parmentier2016}
{Parmentier} V.,  {Fortney} J.~J.,  {Showman} A.~P.,  {Morley} C.,   {Marley} M.~S.,  2016, \mn@doi [\apj] {10.3847/0004-637X/828/1/22}, \href {https://ui.adsabs.harvard.edu/abs/2016ApJ...828...22P} {828, 22}

\bibitem[\protect\citeauthoryear{{Parmentier}, {Showman}  \& {Fortney}}{{Parmentier} et~al.}{2021}]{Parmentier2021MNRAS}
{Parmentier} V.,  {Showman} A.~P.,   {Fortney} J.~J.,  2021, \mn@doi [\mnras] {10.1093/mnras/staa3418}, \href {https://ui.adsabs.harvard.edu/abs/2021MNRAS.501...78P} {501, 78}

\bibitem[\protect\citeauthoryear{{Phillips} et~al.,}{{Phillips} et~al.}{2020}]{Phillips2020A&A...637A..38P}
{Phillips} M.~W.,  et~al., 2020, \mn@doi [\aap] {10.1051/0004-6361/201937381}, \href {https://ui.adsabs.harvard.edu/abs/2020A&A...637A..38P} {637, A38}

\bibitem[\protect\citeauthoryear{{Pierrehumbert} \& {Ding}}{{Pierrehumbert} \& {Ding}}{2016}]{Pierrehumbert2016RSPSA.47260107P}
{Pierrehumbert} R.~T.,  {Ding} F.,  2016, \mn@doi [Proceedings of the Royal Society of London Series A] {10.1098/rspa.2016.0107}, \href {https://ui.adsabs.harvard.edu/abs/2016RSPSA.47260107P} {472, 20160107}

\bibitem[\protect\citeauthoryear{{Pierrehumbert} \& {Hammond}}{{Pierrehumbert} \& {Hammond}}{2019}]{Pierrehumbert2019AnRFM..51..275P}
{Pierrehumbert} R.~T.,  {Hammond} M.,  2019, \mn@doi [Annual Review of Fluid Mechanics] {10.1146/annurev-fluid-010518-040516}, \href {https://ui.adsabs.harvard.edu/abs/2019AnRFM..51..275P} {51, 275}

\bibitem[\protect\citeauthoryear{{Polyansky}, {Kyuberis}, {Zobov}, {Tennyson}, {Yurchenko}  \& {Lodi}}{{Polyansky} et~al.}{2018}]{Polyansky2018}
{Polyansky} O.~L.,  {Kyuberis} A.~A.,  {Zobov} N.~F.,  {Tennyson} J.,  {Yurchenko} S.~N.,   {Lodi} L.,  2018, \mn@doi [\mnras] {10.1093/mnras/sty1877}, \href {http://adsabs.harvard.edu/abs/2018MNRAS.480.2597P} {480, 2597}

\bibitem[\protect\citeauthoryear{{Porco} et~al.,}{{Porco} et~al.}{2003}]{Porco2003Sci...299.1541P}
{Porco} C.~C.,  et~al., 2003, \mn@doi [Science] {10.1126/science.1079462}, \href {https://ui.adsabs.harvard.edu/abs/2003Sci...299.1541P} {299, 1541}

\bibitem[\protect\citeauthoryear{{Powell}, {Zhang}, {Gao}  \& {Parmentier}}{{Powell} et~al.}{2018}]{Powell2018}
{Powell} D.,  {Zhang} X.,  {Gao} P.,   {Parmentier} V.,  2018, \mn@doi [\apj] {10.3847/1538-4357/aac215}, \href {https://ui.adsabs.harvard.edu/abs/2018ApJ...860...18P} {860, 18}

\bibitem[\protect\citeauthoryear{{Rauscher}}{{Rauscher}}{2017}]{Rauscher2017ApJ...846...69R}
{Rauscher} E.,  2017, \mn@doi [\apj] {10.3847/1538-4357/aa81c3}, \href {https://ui.adsabs.harvard.edu/abs/2017ApJ...846...69R} {846, 69}

\bibitem[\protect\citeauthoryear{{Rensen}, {Miguel}, {Zilinskas}, {Louca}, {Woitke}, {Helling}  \& {Herbort}}{{Rensen} et~al.}{2023}]{Rensen2023RemS...15..841R}
{Rensen} F.,  {Miguel} Y.,  {Zilinskas} M.,  {Louca} A.,  {Woitke} P.,  {Helling} C.,   {Herbort} O.,  2023, \mn@doi [Remote Sensing] {10.3390/rs15030841}, \href {https://ui.adsabs.harvard.edu/abs/2023RemS...15..841R} {15, 841}

\bibitem[\protect\citeauthoryear{{Ricker} et~al.,}{{Ricker} et~al.}{2014}]{ricker2014}
{Ricker} G.~R.,  et~al., 2014, in {Oschmann} Jacobus~M. J.,  {Clampin} M.,  {Fazio} G.~G.,   {MacEwen} H.~A.,  eds,  Society of Photo-Optical Instrumentation Engineers (SPIE) Conference Series Vol. 9143, Space Telescopes and Instrumentation 2014: Optical, Infrared, and Millimeter Wave. p. 914320 (\mn@eprint {arXiv} {1406.0151}), \mn@doi{10.1117/12.2063489}

\bibitem[\protect\citeauthoryear{{Roman} \& {Rauscher}}{{Roman} \& {Rauscher}}{2019}]{Roman2019ApJ...872....1R}
{Roman} M.,  {Rauscher} E.,  2019, \mn@doi [\apj] {10.3847/1538-4357/aafdb5}, \href {https://ui.adsabs.harvard.edu/abs/2019ApJ...872....1R} {872, 1}

\bibitem[\protect\citeauthoryear{{Roman}, {Kempton}, {Rauscher}, {Harada}, {Bean}  \& {Stevenson}}{{Roman} et~al.}{2021}]{Roman2021}
{Roman} M.~T.,  {Kempton} E. M.~R.,  {Rauscher} E.,  {Harada} C.~K.,  {Bean} J.~L.,   {Stevenson} K.~B.,  2021, \mn@doi [\apj] {10.3847/1538-4357/abd549}, \href {https://ui.adsabs.harvard.edu/abs/2021ApJ...908..101R} {908, 101}

\bibitem[\protect\citeauthoryear{{Sainsbury-Martinez} et~al.,}{{Sainsbury-Martinez} et~al.}{2023}]{Sainsbury2023MNRAS.524.1316S}
{Sainsbury-Martinez} F.,  et~al., 2023, \mn@doi [\mnras] {10.1093/mnras/stad1905}, \href {https://ui.adsabs.harvard.edu/abs/2023MNRAS.524.1316S} {524, 1316}

\bibitem[\protect\citeauthoryear{{Showman}, {Fortney}, {Lian}, {Marley}, {Freedman}, {Knutson}  \& {Charbonneau}}{{Showman} et~al.}{2009}]{showman2009}
{Showman} A.~P.,  {Fortney} J.~J.,  {Lian} Y.,  {Marley} M.~S.,  {Freedman} R.~S.,  {Knutson} H.~A.,   {Charbonneau} D.,  2009, \mn@doi [\apj] {10.1088/0004-637X/699/1/564}, \href {https://ui.adsabs.harvard.edu/abs/2009ApJ...699..564S} {699, 564}

\bibitem[\protect\citeauthoryear{{Showman}, {Lewis}  \& {Fortney}}{{Showman} et~al.}{2015}]{showman2015}
{Showman} A.~P.,  {Lewis} N.~K.,   {Fortney} J.~J.,  2015, \mn@doi [\apj] {10.1088/0004-637X/801/2/95}, \href {https://ui.adsabs.harvard.edu/abs/2015ApJ...801...95S} {801, 95}

\bibitem[\protect\citeauthoryear{{Sobel}, {Nilsson}  \& {Polvani}}{{Sobel} et~al.}{2001}]{Sobel2001JAtS...58.3650S}
{Sobel} A.~H.,  {Nilsson} J.,   {Polvani} L.~M.,  2001, \mn@doi [Journal of the Atmospheric Sciences] {10.1175/1520-0469(2001)058<3650:TWTGAA>2.0.CO;2}, \href {https://ui.adsabs.harvard.edu/abs/2001JAtS...58.3650S} {58, 3650}

\bibitem[\protect\citeauthoryear{{Spiegel}, {Burrows}  \& {Milsom}}{{Spiegel} et~al.}{2011}]{Spiegel2011ApJ...727...57S}
{Spiegel} D.~S.,  {Burrows} A.,   {Milsom} J.~A.,  2011, \mn@doi [\apj] {10.1088/0004-637X/727/1/57}, \href {https://ui.adsabs.harvard.edu/abs/2011ApJ...727...57S} {727, 57}

\bibitem[\protect\citeauthoryear{{Sromovsky}, {de Pater}, {Fry}, {Hammel}  \& {Marcus}}{{Sromovsky} et~al.}{2015}]{Sromovsky2015Icar..258..192S}
{Sromovsky} L.~A.,  {de Pater} I.,  {Fry} P.~M.,  {Hammel} H.~B.,   {Marcus} P.,  2015, \mn@doi [\icarus] {10.1016/j.icarus.2015.05.029}, \href {https://ui.adsabs.harvard.edu/abs/2015Icar..258..192S} {258, 192}

\bibitem[\protect\citeauthoryear{{Stephan}, {Naoz}  \& {Gaudi}}{{Stephan} et~al.}{2021}]{Stephan2021ApJ...922....4S}
{Stephan} A.~P.,  {Naoz} S.,   {Gaudi} B.~S.,  2021, \mn@doi [\apj] {10.3847/1538-4357/ac22a9}, \href {https://ui.adsabs.harvard.edu/abs/2021ApJ...922....4S} {922, 4}

\bibitem[\protect\citeauthoryear{{Stock}, {Kitzmann}  \& {Patzer}}{{Stock} et~al.}{2022}]{Stock2022MNRAS.517.4070S}
{Stock} J.~W.,  {Kitzmann} D.,   {Patzer} A. B.~C.,  2022, \mn@doi [\mnras] {10.1093/mnras/stac2623}, \href {https://ui.adsabs.harvard.edu/abs/2022MNRAS.517.4070S} {517, 4070}

\bibitem[\protect\citeauthoryear{{Tan}}{{Tan}}{2022}]{Tan2022}
{Tan} X.,  2022, \mn@doi [\mnras] {10.1093/mnras/stac344}, \href {https://ui.adsabs.harvard.edu/abs/2022MNRAS.511.4861T} {511, 4861}

\bibitem[\protect\citeauthoryear{{Tan} \& {Komacek}}{{Tan} \& {Komacek}}{2019}]{Tan&Komacek2019}
{Tan} X.,  {Komacek} T.~D.,  2019, \mn@doi [\apj] {10.3847/1538-4357/ab4a76}, \href {https://ui.adsabs.harvard.edu/abs/2019ApJ...886...26T} {886, 26}

\bibitem[\protect\citeauthoryear{{Tan} \& {Showman}}{{Tan} \& {Showman}}{2019}]{Tan2019Showman}
{Tan} X.,  {Showman} A.~P.,  2019, \mn@doi [\apj] {10.3847/1538-4357/ab0c07}, \href {https://ui.adsabs.harvard.edu/abs/2019ApJ...874..111T} {874, 111}

\bibitem[\protect\citeauthoryear{{Tan} \& {Showman}}{{Tan} \& {Showman}}{2021}]{TanShowman2021b}
{Tan} X.,  {Showman} A.~P.,  2021, \mn@doi [\mnras] {10.1093/mnras/stab097}, \href {https://ui.adsabs.harvard.edu/abs/2021MNRAS.502.2198T} {502, 2198}

\bibitem[\protect\citeauthoryear{{Teinturier} et~al.,}{{Teinturier} et~al.}{2024}]{Teinturier2024A&A...683A.231T}
{Teinturier} L.,  et~al., 2024, \mn@doi [\aap] {10.1051/0004-6361/202347069}, \href {https://ui.adsabs.harvard.edu/abs/2024A&A...683A.231T} {683, A231}

\bibitem[\protect\citeauthoryear{{Toon}, {McKay}, {Ackerman}  \& {Santhanam}}{{Toon} et~al.}{1989}]{toon1989}
{Toon} O.~B.,  {McKay} C.~P.,  {Ackerman} T.~P.,   {Santhanam} K.,  1989, \mn@doi [\jgr] {10.1029/JD094iD13p16287}, \href {https://ui.adsabs.harvard.edu/abs/1989JGR....9416287T} {94, 16287}

\bibitem[\protect\citeauthoryear{{Tsai}, {Lee}  \& {Pierrehumbert}}{{Tsai} et~al.}{2022}]{Tsai2022}
{Tsai} S.-M.,  {Lee} E. K.~H.,   {Pierrehumbert} R.,  2022, \mn@doi [\aap] {10.1051/0004-6361/202142816}, \href {https://ui.adsabs.harvard.edu/abs/2022A&A...664A..82T} {664, A82}

\bibitem[\protect\citeauthoryear{Vallis}{Vallis}{2006}]{vallis2006}
Vallis G.~K.,  2006, Atmospheric and Oceanic Fluid Dynamics: Fundamentals and Large-scale Circulation.
Cambridge University Press, \mn@doi{10.1017/CBO9780511790447}, \url {https://doi.org/10.1017/CBO9780511790447}

\bibitem[\protect\citeauthoryear{Van~Rossum \& Drake~Jr}{Van~Rossum \& Drake~Jr}{1995}]{van1995python}
Van~Rossum G.,  Drake~Jr F.~L.,  1995, Python tutorial.
Centrum voor Wiskunde en Informatica Amsterdam, The Netherlands, \url {https://www.python.org/}

\bibitem[\protect\citeauthoryear{{Vanderburg} et~al.,}{{Vanderburg} et~al.}{2020}]{Vanderburg2020Natur.585..363V}
{Vanderburg} A.,  et~al., 2020, \mn@doi [\nat] {10.1038/s41586-020-2713-y}, \href {https://ui.adsabs.harvard.edu/abs/2020Natur.585..363V} {585, 363}

\bibitem[\protect\citeauthoryear{{Visscher}}{{Visscher}}{2020}]{Visscher2020JGRE..12506526V}
{Visscher} C.,  2020, \mn@doi [Journal of Geophysical Research (Planets)] {10.1029/2020JE006526}, \href {https://ui.adsabs.harvard.edu/abs/2020JGRE..12506526V} {125, e06526}

\bibitem[\protect\citeauthoryear{{Waskom}}{{Waskom}}{2021}]{waskom2021}
{Waskom} M.,  2021, \mn@doi [The Journal of Open Source Software] {10.21105/joss.03021}, \href {https://ui.adsabs.harvard.edu/abs/2021JOSS....6.3021W} {6, 3021}

\bibitem[\protect\citeauthoryear{{Wong}, {Mahaffy}, {Atreya}, {Niemann}  \& {Owen}}{{Wong} et~al.}{2004}]{Wong2004Icar..171..153W}
{Wong} M.~H.,  {Mahaffy} P.~R.,  {Atreya} S.~K.,  {Niemann} H.~B.,   {Owen} T.~C.,  2004, \mn@doi [\icarus] {10.1016/j.icarus.2004.04.010}, \href {https://ui.adsabs.harvard.edu/abs/2004Icar..171..153W} {171, 153}

\bibitem[\protect\citeauthoryear{{Yurchenko}, {Mellor}, {Freedman}  \& {Tennyson}}{{Yurchenko} et~al.}{2020}]{Yurchenko2020}
{Yurchenko} S.~N.,  {Mellor} T.~M.,  {Freedman} R.~S.,   {Tennyson} J.,  2020, \mn@doi [\mnras] {10.1093/mnras/staa1874}, \href {https://ui.adsabs.harvard.edu/abs/2020MNRAS.496.5282Y} {496, 5282}

\bibitem[\protect\citeauthoryear{{Zhan}, {Koll}  \& {Ding}}{{Zhan} et~al.}{2024}]{Zhan2024ApJ...971..125Z}
{Zhan} R.,  {Koll} D. D.~B.,   {Ding} F.,  2024, \mn@doi [\apj] {10.3847/1538-4357/ad54c1}, \href {https://ui.adsabs.harvard.edu/abs/2024ApJ...971..125Z} {971, 125}

\bibitem[\protect\citeauthoryear{{Zuckerman}, {Melis}, {Klein}, {Koester}  \& {Jura}}{{Zuckerman} et~al.}{2010}]{Zuckerman2010ApJ...722..725Z}
{Zuckerman} B.,  {Melis} C.,  {Klein} B.,  {Koester} D.,   {Jura} M.,  2010, \mn@doi [\apj] {10.1088/0004-637X/722/1/725}, \href {https://ui.adsabs.harvard.edu/abs/2010ApJ...722..725Z} {722, 725}

\bibitem[\protect\citeauthoryear{pandas~development team}{pandas~development team}{2020}]{reback2020pandas}
pandas~development team T.,  2020, pandas-dev/pandas: Pandas, \mn@doi{10.5281/zenodo.3509134}, \url {https://doi.org/10.5281/zenodo.3509134}

\makeatother
\end{thebibliography}




\appendix

\section{Weak temperature gradient theory}
\label{app:WTG}

\citet{Pierrehumbert2019AnRFM..51..275P} described the weak temperature gradient theory which is summarised in brief in this appendix. There is a weak temperature gradient (WTG) if the rotation rate is so low that the Coriolis forces become negligible. Furthermore, if the frictional forces and non-linearities in the momentum equation are as well negligible, then only the pressure gradient remains in the steady-state horizontal momentum equation, which must therefore vanish. In this case, there are vanishing pressure gradients at large scales and consequently vanishing horizontal temperature gradients. In the case, the Coriolis force, friction, and non-linearities are not zero, the WTGs are weak \citep{Pierrehumbert2019AnRFM..51..275P}.

The WTG relates the planets spin at a specific ratio to the planetary rotation rate, $\Omega$ [rad s$^{-1}$]  \citep{Pierrehumbert2019AnRFM..51..275P}. The WTG behaviour starts at the critical length scale that is the Rossby radius of deformation, $L_d$ [m] \citep{vallis2006}. The global WTG behaviour includes a relevant radius of deformation, $c_0/\Omega$, where $c_0$ [m s$^{-1}$] is the fastest gravity wave speed for non-rotating flow. $L_d$ describes how far fast gravity waves can travel without the effects of Coriolis and geostrophically adjust. Thus, $L_d$ is a characteristic spatial scale that avoids the effect of local heating. Anyhow, the WTG behaviour also relates to the unperturbed layer depth. In shallow water theory, the unperturbed layer depth is  simply described  by the scale height, $H$ [m]. In a compressible atmosphere, the $H$ is defined \citep{Pierrehumbert2019AnRFM..51..275P} as
\begin{equation}
H = \frac{R_dT}{g},
 \label{eq:scale_height}
\end{equation}
where $T$ [K] is the temperature of the gas, $g$ [m s$^{-2}$] the gravity, and $R_{d}$ [J kg$^{-1}$ K$^{-1}$] the specific gas constant (here the constant in the dynamical core). The $c_0$ is \citep{Pierrehumbert2019AnRFM..51..275P} then
\begin{equation}
c_0 = \sqrt{gH}= \sqrt{R_dT}.  
 \label{eq:c_0}
\end{equation}
Considering shallow water dynamics, we can calculate the 
layer depth as an adiabatic layer with potential temperature, $\theta_0$ [K]. We computed $\theta$ with the local temperature and took the mean to get $\theta_0$. Then, $c_0$ becomes \citep{Pierrehumbert2019AnRFM..51..275P}
\begin{equation}
c_0 = \sqrt{R_d\theta_0}.
 \label{eq:c_2}
\end{equation}
Therefore, we compute the $c_0$ as the isothermal speed of sound, that deviates from the adiabatic sound speed by a factor of unity.
Finally, the WTG parameter, $\Lambda$, is defined \citep{Pierrehumbert2019AnRFM..51..275P} as
\begin{equation}
\Lambda = \frac{L_d}{a} = \frac{c_0}{\Omega a} = \frac{\sqrt{R_d\theta_0}}{\Omega a}, 
 \label{eq:WTG}
\end{equation}
where a [m] is the radius of the planet.

$\Lambda$ may describe the dynamical behaviour of the primitive equation set in the dynamical core of Exo-FMS GCM when applied to WD-1856b. In reality, the shallow water theory is not suitable for the large depth of this planet. Therefore, $\Lambda$ describes the dynamics created by the GCM rather than the real behaviour.

For $\Lambda \gg 1$, global WTG behaviour dominates. For $\Lambda \ll 1$, strong temperature gradients evolve across the planet due to a strong contrast in geographical heating. For $\Lambda \sim 1$, WTG behaviour occurs near the equator, but strong temperature gradients evolve at higher latitudes. The Earth has this condition ($\Lambda \simeq 0.6$), however, non-WTG behaviour can still occur for $\Lambda \gg 1$ due to non-linearities \citep{Pierrehumbert2019AnRFM..51..275P,Charney1963JAtS...20..607C,Sobel2001JAtS...58.3650S,Pierrehumbert2016RSPSA.47260107P}.

\bsp	
\label{lastpage}
\end{document}